\documentclass{aastex61}
\usepackage{graphicx,natbib}
\usepackage[percent]{overpic}
\usepackage{graphicx,subfigure,amsmath}
\newcommand\gsim{\,\lower3pt\hbox{$\sim$}\llap{\raise2pt\hbox{$>$}}\,}
\newcommand\lsim{\,\lower3pt\hbox{$\sim$}\llap{\raise2pt\hbox{$<$}}\,}

\shorttitle{prominence eruption}
\shortauthors{Fan}

\begin{document}

\title{MHD simulation of prominence eruption}

\correspondingauthor{Yuhong Fan}
\email{yfan@ucar.edu}

\author{Yuhong Fan}
\affil{High Altitude Observatory, National Center for Atmospheric Research, \\
3080 Center Green Drive, Boulder, CO 80301, USA}

\begin{abstract}
We carry out magnetohydrodynamic (MHD) simulations of the quasi-static evolution and
eruption of a twisted coronal flux rope under a coronal streamer built up by an imposed
flux emergence at the lower boundary.
The MHD model
incorporates a simple empirical coronal heating, optically thin radiative
cooling, and field aligned thermal conduction, and thus allows the
formation of prominence condensations.
We find that during the quasi-static evolution,
prominence/filament condensations of an elongated, sigmoid morphology form
in the dips of the significantly twisted field lines of the emerged flux rope due
to run-away radiative cooling.
A prominence cavity also forms surrounding the prominence, which is best observed
above the limb with the line-of-sight nearly along the length of the
flux rope, as shown by synthetic SDO/AIA EUV images.
The magnetic field supporting the prominence is significantly non-force-free
despite the low plasma-$\beta$.
By comparing with a simulation that suppresses prominence formation, we find
that the prominence weight is dynamically important and
can suppress the onset of the kink instability and hold the flux rope in
equilibrium for a significantly long time, until draining
of the prominence plasma develops and lightens the prominence weight.
The flux rope eventually develops the kink instability and erupts, producing a
prominence eruption. The synthetic AIA 304 {\AA} images show that the
prominence is lifted up into an erupting loop, exhibiting helical features
along the loop and substantial draining along the loop legs, as often
seen in observations.
\end{abstract}

\keywords{magnetohydrodynamics(MHD) --- methods: numerical --- Sun: corona --- Sun: coronal mass ejections (CMEs) --- Sun: filaments, prominences}

\section{Introduction}
Observations have shown a close association between prominence/filament
eruptions and coronal mass ejections (CMEs)
\citep[e.g.][]{Munro:etal:1979,Webb:Hundhausen:1987}.
This indicates that understanding the hosting magnetic field structures
capable of supporting the prominence/filament are important for understanding
the origin of CMEs.
One type of model for the hosting magnetic structure
is a magnetic flux rope with helical
field lines twisting about its central axis, with the cool and dense prominence
plasma suspended in the dips of the field lines in the much hotter and
rarefied corona \citep[e.g.][]{Low:2001,Gibson:2015}.
Many MHD simulations have been carried out to study the mechanisms, such as
the helical kink instability and the torus instability,
for the destabilization and eruption of a coronal flux rope
and successfully achieved the eruptive behavior \citep[e.g.][]{
Toeroek:Kliem:2005,Toeroek:Kliem:2007,Toeroek:etal:2011,Aulanier:etal:2010,
Fan:Gibson:2007,Fan:2010,Fan:2012, Chatterjee:Fan:2013,Amari:etal:2014}.
These simulations have generally used highly
simplified thermodynamics, e.g. zero-plasma $\beta$, isothermal, or
a polytropic gas with lowered adiabatic index $\gamma$, without the
possibility for the formation of the cool prominence/filament condensations.
Thus the possible dynamic role the prominence mass plays on the
stability and eruptive properties of the coronal flux
ropes \citep[e.g.][]{Low:1996} have not been well modeled in MHD
simulations of CMEs, despite observational indications of its
importance \citep[e.g.][]{Jenkins:etal:2017}.

Increasingly, 3D MHD simulations of CMEs and CME source regions are
conducted with more realistic treatment of the thermodynamics to allow
for a more direct comparison with coronal multi-wavelength
observations \citep[e.g.][]{Downs:etal:2012,Toeroek:etal:2018}
and to model the formation of prominences/filaments
\citep[e.g.][]{Xia:etal:2014,Xia:Keppens:2016a,Fan:2017}.
\citet{Xia:etal:2014} and \citet{Xia:Keppens:2016a} have carried out
the first 3D MHD simulations of the formation of a prominence in a stable
equilibrium coronal flux rope, by incorporating the important non-adiabatic
effects of an empirical coronal heating, optically thin radiative losses,
and field aligned thermal conduction. They have also used an adaptive
grid to resolve the fine-scale internal dynamics of the prominence.
Their model obtained a prominence-cavity system with
the prominence containing fine-scale, highly dynamic fragments, which
reproduced many observed features seen in SDO/AIA observations.
Recently, \citet{Fan:2017} has carried out a first 3D MHD simulation of
a prominence-carrying coronal flux rope that
transitions from quasi-equilibrium to eruption,
with the thermodynamics treatment incorporating the non-adiabatic
effects of a simple empirical coronal heating, optically thin radiative
losses, and field-aligned thermal conduction.
In that simulation a significantly twisted, long coronal flux rope
builds up under a pre-existing coronal streamer by an imposed flux emergence
at the lower boundary.
It is found that during the quasi-static phase of the evolution of the emerged
flux rope, cool prominence condensations form in the dips of the twisted
field lines due to in-situ radiative instability driven by the optically
thin radiative cooling.  Subsequently, the flux rope erupts as it develops
the kink instability, and a prominence eruption is produced with substantial
draining of the prominence plasma.
That simulation also shows that once the prominence is formed, the magnetic
field supporting the prominence becomes significantly non-force-free,
despite the fact that the entire flux rope has low plasma-$\beta$. 
This suggests that the weight of the prominence mass may be dynamically
important.

In this paper we use the same numerical model described in
\citet{Fan:2017} to carry out further simulations of the evolution of a
significantly twisted coronal flux rope under a coronal streamer built
up by an imposed flux emergence at the lower boundary.
By changing the specification of the lower boundary coronal base pressure,
we obtained the formation of a more massive and extended prominence with
more dips forming prominence condensations in the emerged flux rope. As a
result we also found a more continuous and extended low density region
surrounding the elongated prominence/filament in the flux rope, producing
a prominence-cavity system in the synthetic AIA EUV images with the
flux rope observed above the limb viewed along a line-of-sight nearly
along the length of the flux rope.
By comparing with a corresponding simulation where we
suppress the formation of prominence condensations, we found that the
weight of the prominence mass can suppress or delay the onset of the kink
instability of an otherwise nearly force-free flux rope, until
significant draining of the prominence develops.

\section{Model Description}
\label{sec:model}

\begin{figure}[htb!]
\centering
\includegraphics[width=0.6\textwidth]{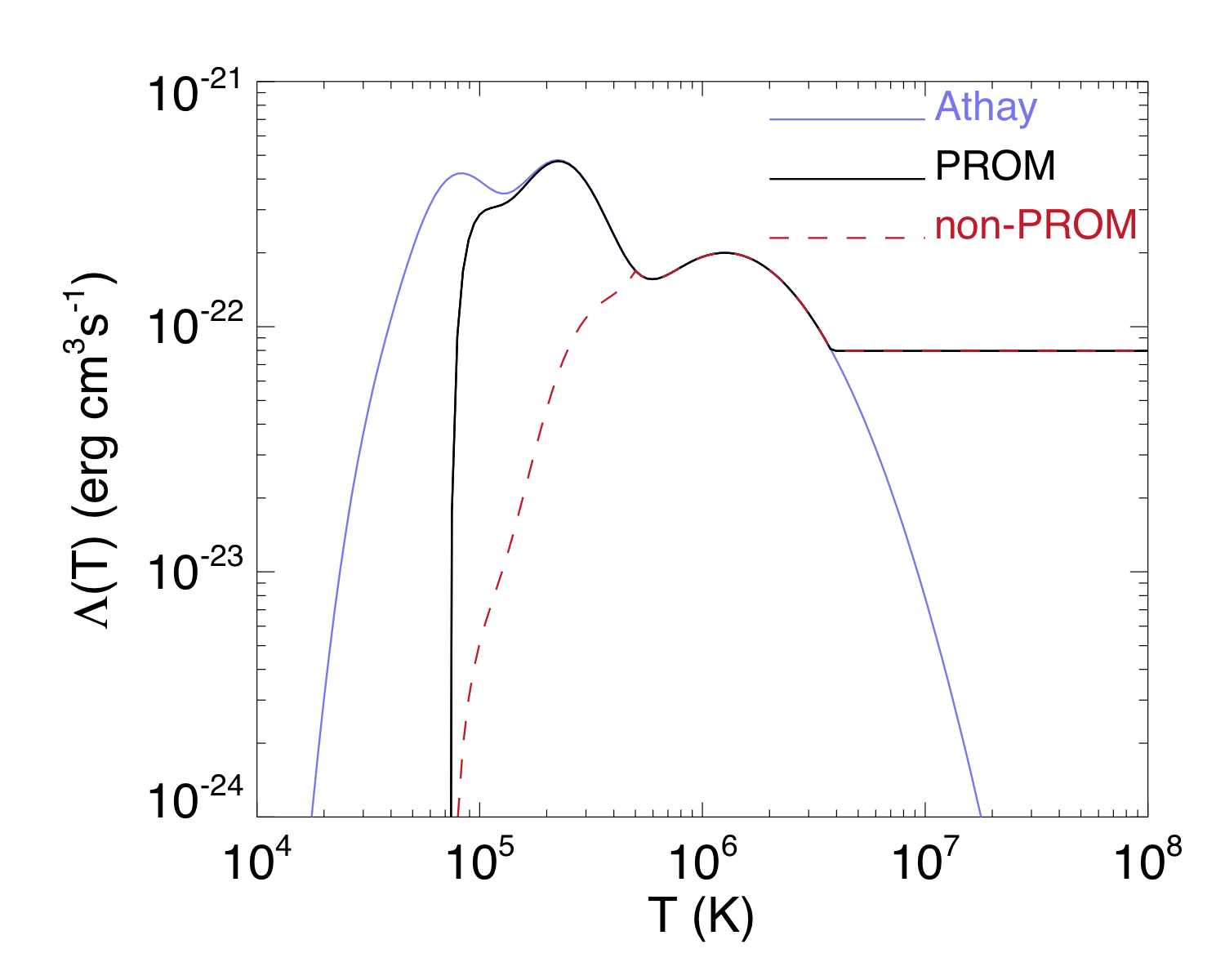}
\caption{The radiative loss function $\Lambda (T)$ defined in equation (13)
in F17. The blue curve is the form given in \citet{Athay:1986}.
The modified $\Lambda (T)$ used for the PROM case (black
curve) is the same as that used in F17 (see section 2 of F17 for the
description of the modifications and the reasons), and the further
modified function (red dashed curve) is used for the non-PROM case,
where the radiative loss is further reduced for
$T \leq 5 \times 10^5$ K compared to the PROM case.}
\label{fig:radloss}
\end{figure}

The numerical simulations of this work use the ``Magnetic Flux Eruption''
(MFE) code to solve the same set of semi-relativistic MHD equations in
spherical geometry as described in detail in \citet[][hereafter F17]{Fan:2017}.
The thermodynamics treatment of the simulations explicitly takes into
account the non-adiabatic effects
of an empirical coronal heating (that depends on height only),
optically thin radiative cooling, and field aligned heat conduction.
The simulation domain is in the corona, ignoring the photosphere and
chromosphere layers, with the lower boundary temperature and density
set at the base of the corona.
The detailed simulation setups, including the equations solved,
the numerical algorithms used,
and the initial state and the boundary conditions, are all as
described in F17, and the readers are referred to section 2 and section 3.1
in that paper for these descriptions. The specific changes made for the
simulations of this paper are described below.
All simulations in this paper use the set up
(in regard to the simulation domain, initialization of the initial helmet
streamer solution, and the boundary conditions) of the
``WS-L'' case described in F17
(where ``WS-L'' denotes Wide Streamer and
Long curvature radius of the driving emerging torus in F17).
As a brief summary of the set up,
we first initialize a 2D quasi-steady solution
of a coronal steamer with an ambient solar wind in a spherical wedge domain
with $r \in [R_s, 11.47 R_s]$,
$\theta \in [75^{\circ}, 105^{\circ}]$,
and $\phi \in [-75^{\circ}, 75^{\circ}]$, where $R_s$ is the solar radius.
Into the dome of the
streamer we drive the slow emergence of a portion of a twisted magnetic
torus at the lower boundary (see equations (19)-(22) and associated
description in F17), so that a
long twisted flux rope is built up
quasi-statically under the steamer, and we study its resulting dynamic
evolution in the corona.
One change made for the simulations in this paper is that we have used
different values for
the proportionality parameter $C$ in equation (18) in F17. That equation
determines the (varying) lower boundary pressure in response to the
downward heat conduction flux, to crudely represent the effect of
chromospheric evaporation as described in F17.
We have carried out two simulations in this paper, one develops
prominence condensations (hereafter referred to as the ``PROM'' case),
and the other does not (hereafter referred to as the ``non-PROM'' case) for
comparison.
For the PROM case, we have used a $C$ value that is increased to about
1.5 times the value used in F17.
This increases the coronal base pressure at the lower boundary
given the same downward heat condition, and it results in the formation
of a more massive and extended prominence in the emerged
coronal flux rope compared to the WS-L case in F17, as described in
section \ref{sec:results} below.
For the non-PROM case, we have used the same $C$ value as that in F17.
But we have modified the radiative loss function ($\Lambda (T)$ in  F17)
as shown in the red dashed curve in Figure \ref{fig:radloss} to reduce
cooling for temperature $T$ below $5 \times 10^5$ K compared to that
used for the PROM case (black curve, which is the same as the one used
in WS-L case in F17).
We have also enhanced thermal
conduction for $T$ below $T_0 = 5 \times 10^5$ K in the non-PROM case:
for the thermal conduction heat flux given by equation (10) in F17,
the thermal conductivity $\kappa_0 T^{5/2}$ is set to a constant value of
$\kappa_0 {T_0}^{5/2}$ for $T < T_0$.  With these changes in the
non-PROM case we are able to suppress the development of runaway
radiative cooling and
prevent the formation of prominence condensations with temperatures
below $10^5$ K.
Finally, the field strength for the emerging magnetic torus used
for driving the lower boundary flux
emergence, $B_t a/R'$ in Table 1 of F17, is set to 103 G
for both the PROM and non-PROM simulations
(instead of the 100 G for the WS-L case in F17).

\section{Simulation results}
\label{sec:results}

\subsection{Simulations with and without the prominence: the dynamical effect of the
prominence mass}
\begin{figure}[htb!]
\centering
\includegraphics[width=0.3\textwidth]{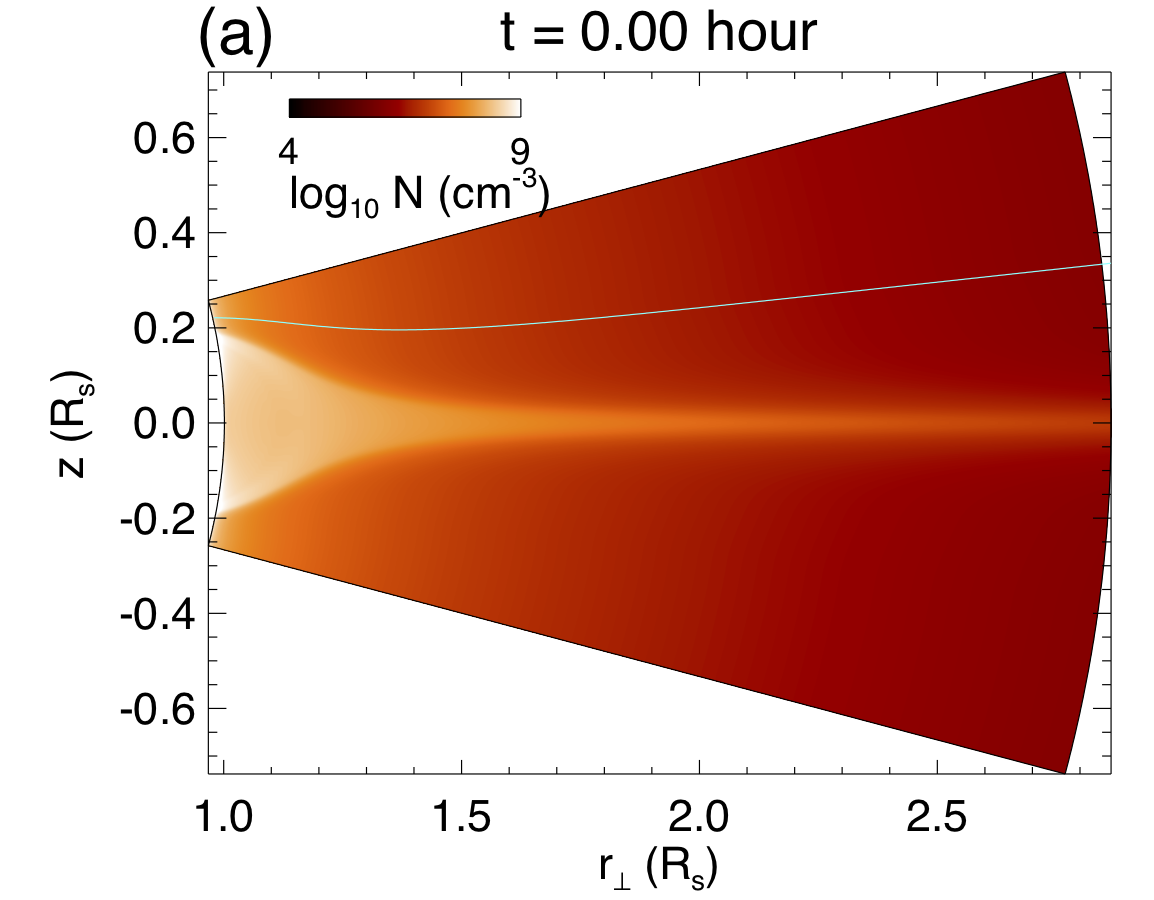}
\includegraphics[width=0.3\textwidth]{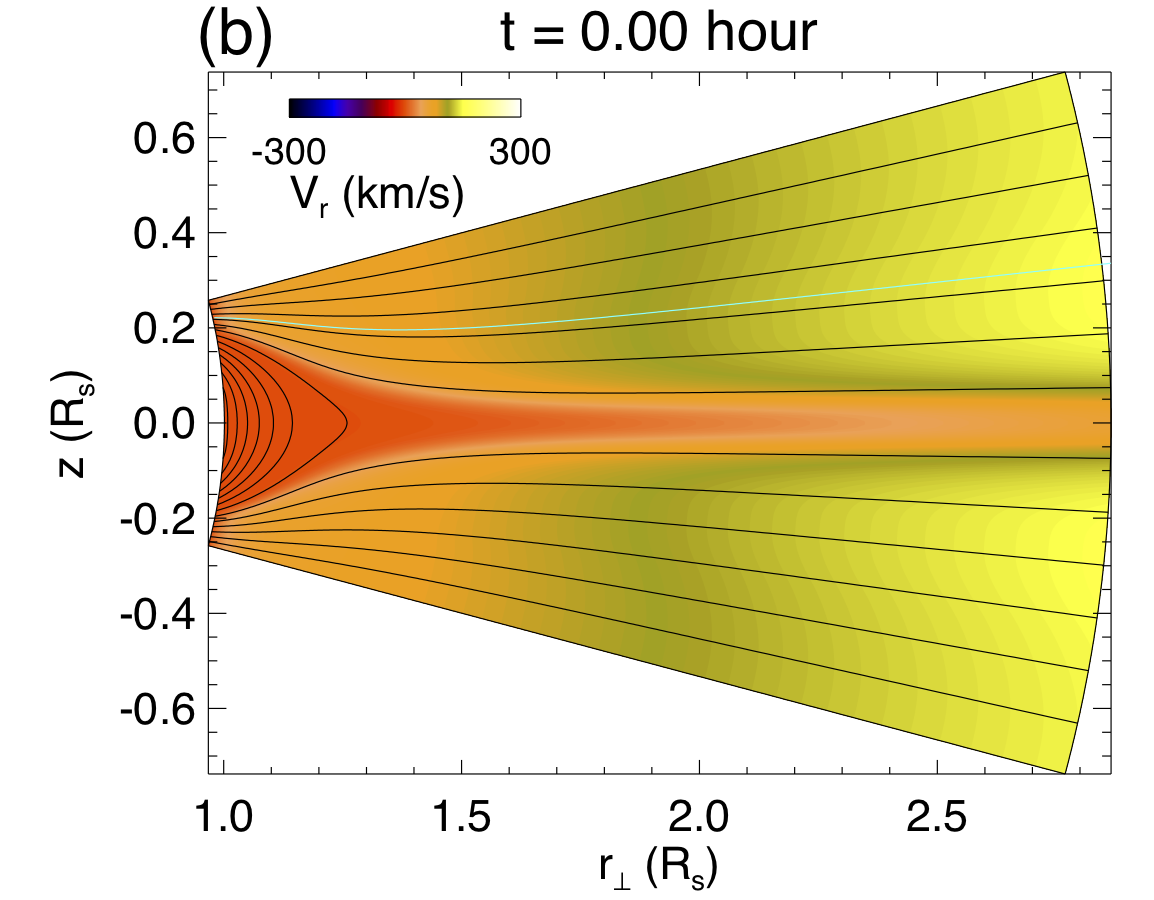} \\
\includegraphics[width=0.3\textwidth]{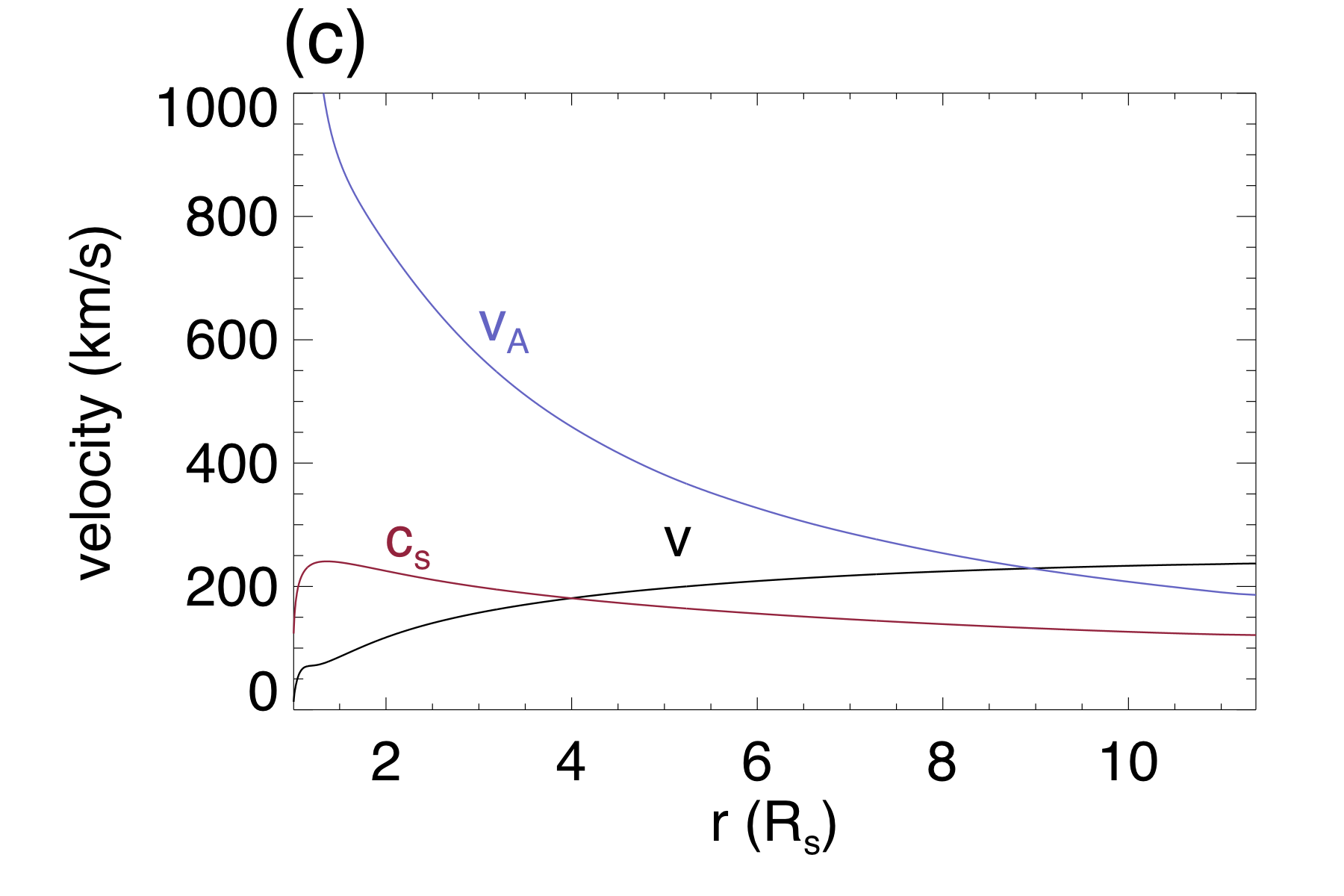}
\includegraphics[width=0.3\textwidth]{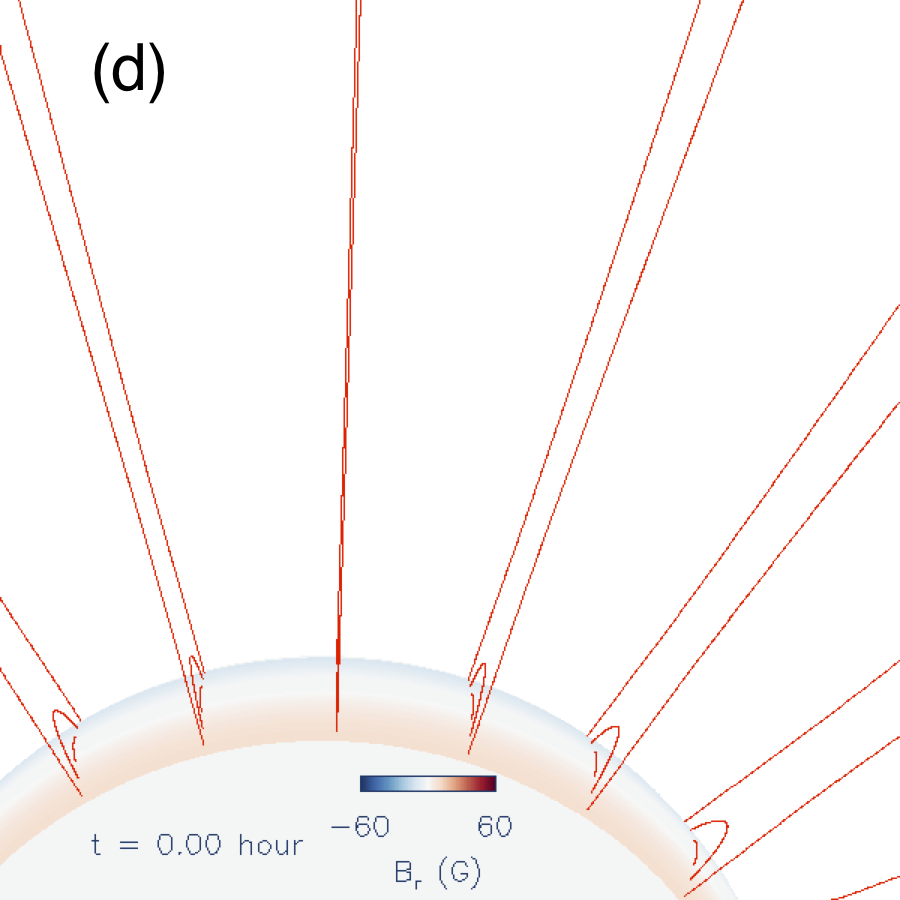}
\caption{The initial streamer solution for the PROM simulation:
(a) the density in the meridional cross-section;
(b) the radial velocity in the meridional cross-section over plotted with
the magnetic field lines; (c) the parallel velocity $V$,
the Alfv\'en speed $V_a$, and the sound speed $C_s$, along an open field line
(the green line shown in panel (a) and (b));
(d) a 3D view of selected initial field lines in the simulation domain
with the lower boundary color indicating the normal
magnetic field $B_r$.}
\label{fig:WS}
\end{figure}

For the initial state of the PROM simulation, we first initialize a 2D quasi-steady
solution of a coronal streamer with a background solar wind, following the procedure
described in section 3.1 of F17 for setting up the WS (wide streamer) solution. 
The relaxed initial streamer solution is shown in Figure \ref{fig:WS}.
Although here we have used a different constant $C$ value (1.5 times the value used
in F17) for setting the lower boundary pressure in equation (18) in F17,
the resulting initial streamer solution remains very
close to the initial WS streamer solution shown in Figure 3 of F17.
As can be seen in Figure \ref{fig:WS}(a)(b), the initial state is a helmet
streamer with a denser dome of closed magnetic field in approximate static equilibrium,
surrounded by an ambient lower density open field region with an outflow.
The outflow speed, the Alfv\'en speed, and the sound speed along an open field line
(the green field line in Figures \ref{fig:WS}(a)(b)) as shown in Figure \ref{fig:WS}(c)
are all very close to those of the WS solution in F17.
Although the increased $C$ value tends to drive a higher coronal base pressure, the
more mass flowing into the corona enhances cooling and in turn reduces downward heat
conduction, thus as a net outcome the base
pressure increase for the initial helmet steamer solution is small
($\lsim 10$\%)
and the overall initial state remains very close to the WS initial state in F17.

\begin{figure}[htb!]
\centering
\includegraphics[width=0.84\textwidth]{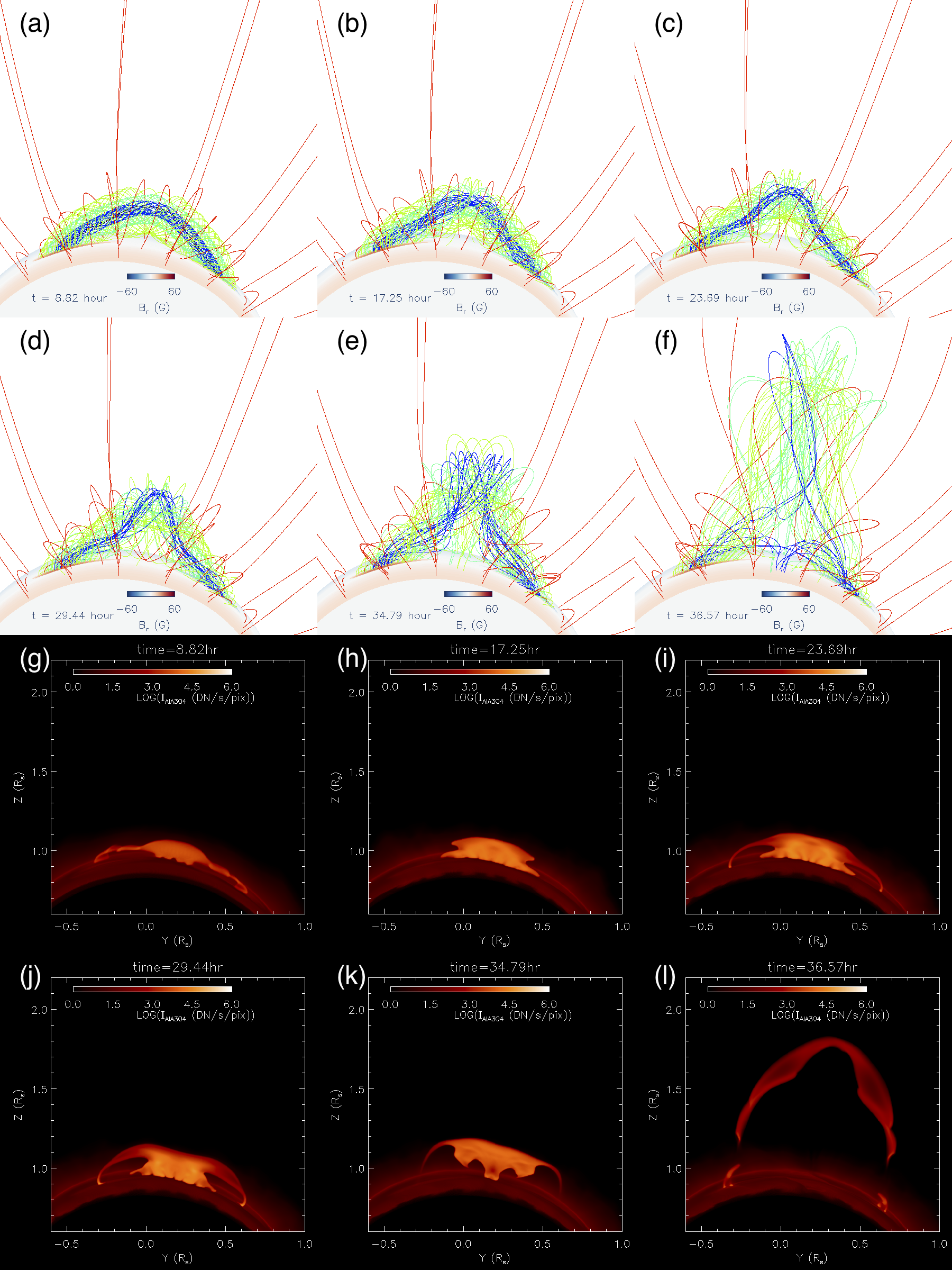}
\caption{The 3D evolution obtained from the PROM simulation. Panels (a)(b)(c)(d)(e)(f)
show snapshots of the 3D magnetic field lines. The lower boundary surface
is colored with $B_r$. The field lines are colored based on the original
flux surfaces of the driving emerging magnetic torus as described in the text.
Panels (g)(h)(i)(j)(k)(l) show the corresponding synthetic SDO/AIA 304 {\AA}
channel emission images as viewed from the same line of sight as the field
line images.  A movie corresponding to this figure
showing the evolution of the 3D magnetic field and the synthetic AIA 304 {\AA}
emission, from $t=6$ hour to about $t=40$ hour is available in the online
version of the paper.}
\label{fig:prom_evol}
\end{figure}

\begin{figure}[htb!]
\centering
\includegraphics[width=0.5\textwidth]{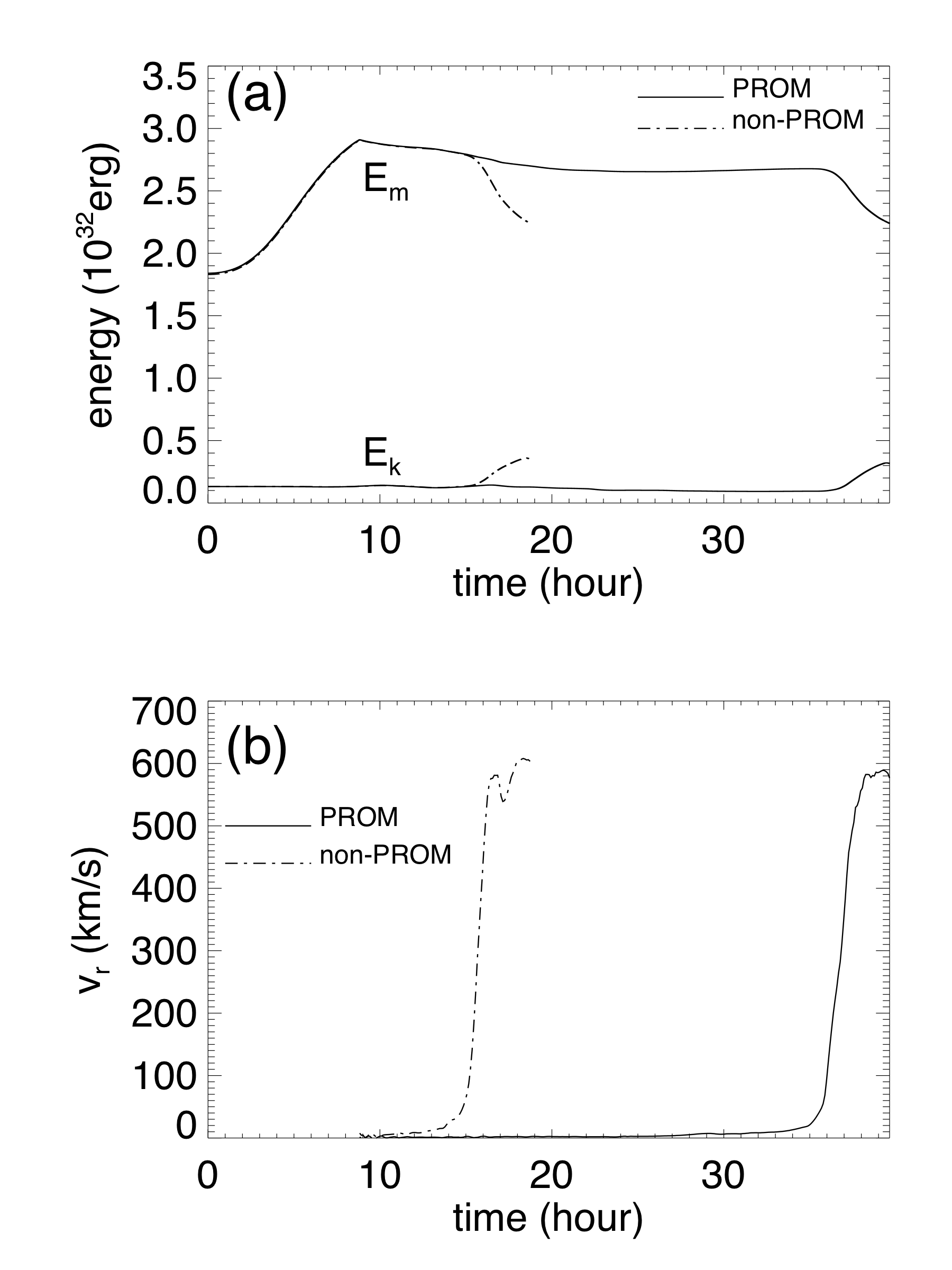}
\caption{Panel (a) shows the temporal evolution of the total magnetic energy
$E_m$ and total kinetic energy $E_k$ for the PROM case (solid curves) and
the non-PROM case (dash-dotted curve).  Panel (b) shows the temporal
evolution of the rise velocity $v_r$ tracked at the
apex of the axial field line of the emerged flux rope for the PROM case
(solid curve) and the non-PROM case (dash-dotted curve)}
\label{fig:emekvr}
\end{figure}

Into the initial helmet dome, we drive the emergence of a magnetic torus
at the lower boundary in the same way as in the WS-L simulation described in
section 3.1 of F17.
The resulting evolution produced by the PROM simulation is shown in Figure
\ref{fig:prom_evol} and the associated movie in the on-line version.
The top panels ((a)-(f)) show a sequence of snapshots of the 3D magnetic
field lines. The field lines are colored as follows (same way as was done
for Figure 5 of F17). A set of field lines traced from a fixed set of
footpoints in the initial bipolar bands outside of the emerging
flux region are colored red. For tracing the field lines
from the emerging flux region, we track the
footpoints on the lower boundary that connect to a fixed set of field
lines of the subsurface emerging torus and color the field lines
(green and blue field lines, and the black axial field line)
based on the flux surfaces of the subsurface
torus the field lines are on. 
The lower panels ((g)-(l)) of Figure \ref{fig:prom_evol} show
synthetic SDO/AIA 304 {\AA} channel emission images corresponding
to each of the 3D field snapshots in the upper panels.  The
synthetic SDO/AIA 304 {\AA} channel emission images
are computed by integration along the line-of-sight (with the same
viewing angle as that for the upper field line snapshots) through the
simulation domain using equation (23) in F17.
It can be seen from the AIA 304 {\AA} emission (which has a peak
temperature response at about $8 \times 10^4$ K), cool prominence
plasma condensations of elongated morphology develop in the
emerged flux rope in the corona.
As described in F17, the formation of the prominence is
due to the onset of radiative instability or runaway
radiative cooling of the plasma in the dips of the emerged flux rope
field lines. We find the formation of a more extended
and massive prominence in the PROM simulation compared to the WS-L case
in F17, as can be seen by comparing the AIA 304 {\AA} emission images
in Figure \ref{fig:prom_evol} with those in Figure 10 of F17.
Figure \ref{fig:emekvr} shows the temporal evolution of the total
magnetic energy $E_m$ and total kinetic energy $E_k$ (panel (a)),
and the rise velocity tracked at the apex of the axial field line
of the flux rope (panel (b)), from the PROM simulation (solid curves).
The ``axial field line'' refers to the field line that is
traced from the footpoints on the lower boundary that connect
to the axis of the driving emerging torus.
The $v_r$ in Figure \ref{fig:emekvr}(b) is evaluated at the apex
position of this axial field line.
The axial field line reconnects during the onset of eruption,
after which we continue to track the Lagrangian evolution of
the the apex plasma element based on its velocity.
Similar to the WS-L simulation in F17, we see
in Figure \ref{fig:emekvr}(a) that the total magnetic energy increases
from $t=0$ to $8.82$ hour as a coronal flux rope
is built up (Figure \ref{fig:prom_evol}(a)) due to the
driving flux emergence at the lower boundary.
The emergence is stopped at $t=8.82$ hour, when the total field line
twist about the axis of the emerged flux rope reaches 1.83 winds,
which is above the critical twist (1.25 winds) for
the development of the kink instability derived for a line-tied,
uniformly twisted cylindrical force-free flux tube
\citep{Hood:Priest:1981}.
After the emergence is stopped, the coronal flux rope is found to
undergo a long quasi-static phase (from $t=8.82$ to about $36$ hours),
during which the rope remains confined under the streamer with close to zero
rise velocity
and with the total magnetic energy slowly declining for most of this
quasi-static phase (see Figures \ref{fig:prom_evol}(a)(b)(c)
and solid curves in Figure \ref{fig:emekvr}).
Towards the later part of the quasi-static phase (after about $t=23$ hour),
we begin to see draining of the prominence mass in multiple branches
(see Figures \ref{fig:prom_evol}(i)(j)(k) and the online movie), and
the flux rope starts a slow rise as it becomes significantly kinked
(see Figures \ref{fig:prom_evol}(c)(d)(e)). Eventually the central
protrusion rises to a height where it can no longer be
confined and develops a hernia-like ejective eruption
(see Figures \ref{fig:prom_evol}(e)(f)) at about $t=36$ hours,
when $E_m$ shows a sudden decrease, $E_k$ shows a sudden
increase, and the rise speed $v_r$ shows a significant acceleration
(see solid curves in Figure \ref{fig:emekvr}).
As can be seen from Figures \ref{fig:prom_evol}(k)(l) and the movie,
the prominence is lifted up into an erupting loop, with
substantial draining along the legs of the erupting loop,
and the draining prominence exhibits helical features as are often observed.
The quasi-static phase after the emergence is stopped and the
eventual hernia like eruption of the flux rope are qualitatively similar
to the evolution in the WS-L case in F17.
However in the present PROM simulation the quasi-static phase
lasts about 3 times longer, about 27 hours, which is about 200 times the
Alfv\'en transit time $\tau_A = 0.137$ hours of the flux rope axis, and
the development of the kink motion of the flux rope is significantly delayed.
The long time scale compared to $\tau_A $ indicates
the flux rope remains close to equilibrium during the quasi-static phase.
The present
flux rope is kept in equilibrium much longer because of the more
massive and extended prominence condensations that form in
the PROM simulation.

\begin{figure}[htb!]
\centering
\includegraphics[width=0.84\textwidth]{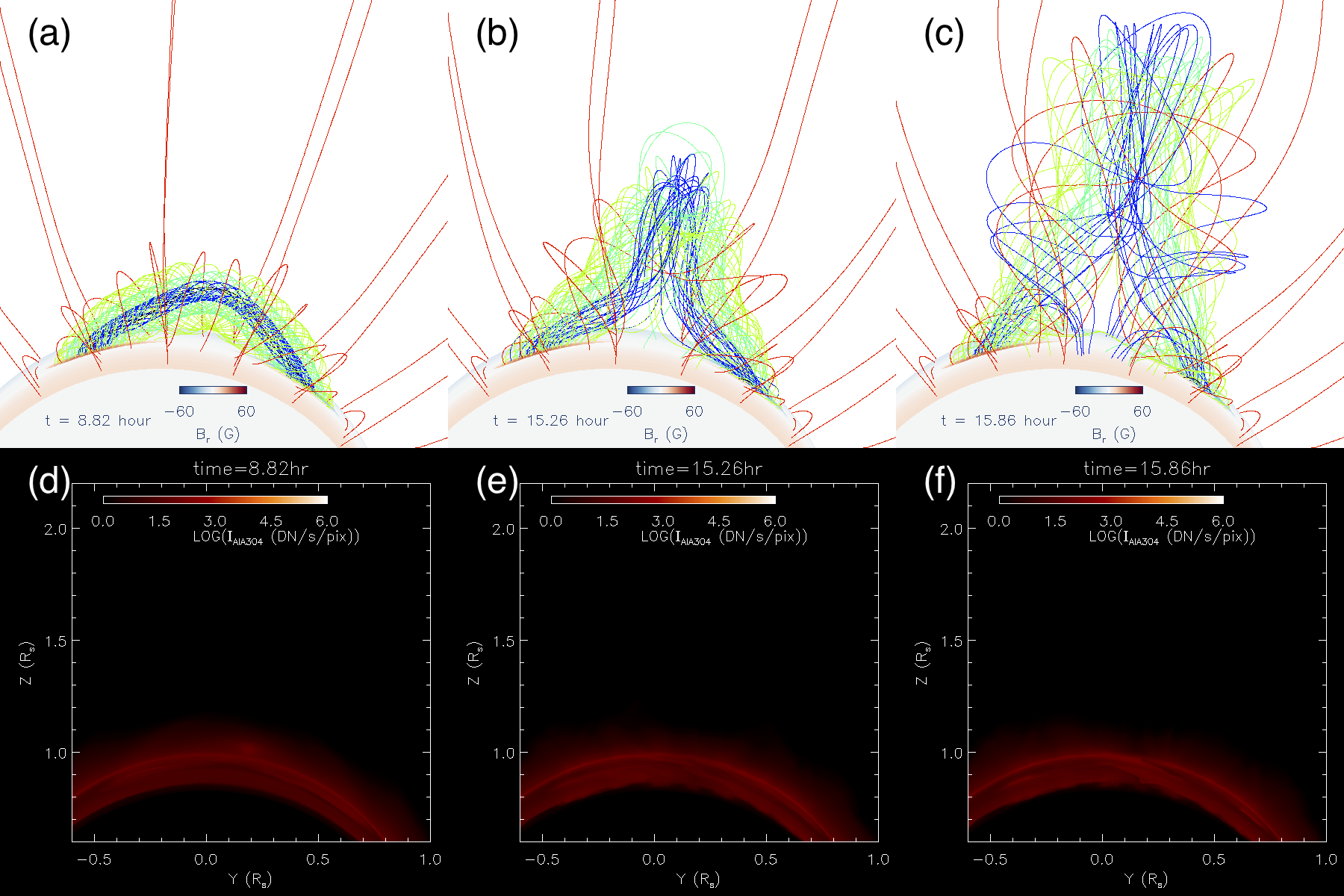}
\caption{Same as Figure \ref{fig:prom_evol} but with fewer snapshots
showing the 3D evolution obtained from the non-PROM simulation.
A movie corresponding to this figure
showing the evolution of the 3D magnetic field and the synthetic AIA 304 {\AA}
emission from $t=6$ hours to about $t=22$ hours is available in the
online version of the paper.}
\label{fig:nonprom_evol}
\end{figure}

To study the effect of the prominence mass, we have also carried out the
non-PROM simulation which suppresses the development of the
cool prominence condensations (by modifying the radiative
cooling and thermal conduction as described in section \ref{sec:model}).
The resulting evolution from the non-PROM simulation is shown in
Figure \ref{fig:emekvr} (dash-dotted curves) and
Figure \ref{fig:nonprom_evol} (with an associated movie in the on-line version).
As is in the case of the PROM simulation, the flux emergence is stopped at
$t=8.82$ hours when a significantly twisted coronal flux rope has been
built up (see Figure \ref{fig:nonprom_evol}(a))
with the total field line twist reaching 1.83 winds about
the anchored axial field line.
However, no significant cool prominence condensation forms in this
case throughout the evolution as can
be seen from the AIA images in Figures \ref{fig:nonprom_evol}(d)(e)(f) and
the movie.
As a result, the development of the kink instability sets in much sooner
as can be seen in Figure \ref{fig:nonprom_evol}(b), where the flux rope has become
significantly kinked and its central portion protruding to a significantly
higher height compared to the prominence carrying flux rope shown in
Figure \ref{fig:prom_evol}(b)
at an even later instant in time (about 2 hours later),
for which there is still little development of the kink motion.
The flux rope in the non-PROM case develops a hernia like eruption at
about $t=15.5 $ hours
as can be seen in Figures \ref{fig:nonprom_evol}(b)(c)
and also in Figure \ref{fig:emekvr} (dash-dotted lines), where $E_m$
undergoes a significant decrease, $E_k$ a significant increase, and the
rise velocity $v_r$ a significant acceleration at about
$t=15.5$ hours,
far sooner than the onset of eruption for the PROM case.
It appears that the formation of the massive prominence condensation
has delayed the growth of the kink instability and the onset of eruption
for a substantial time
period, about 20 hours or $146 \tau_{A}$, compared to the case without the
prominence formation (compare the two curves in
Figure \ref{fig:emekvr}(b)).

\begin{figure}[hbt!]
\centering
\includegraphics[width=0.5\textwidth]{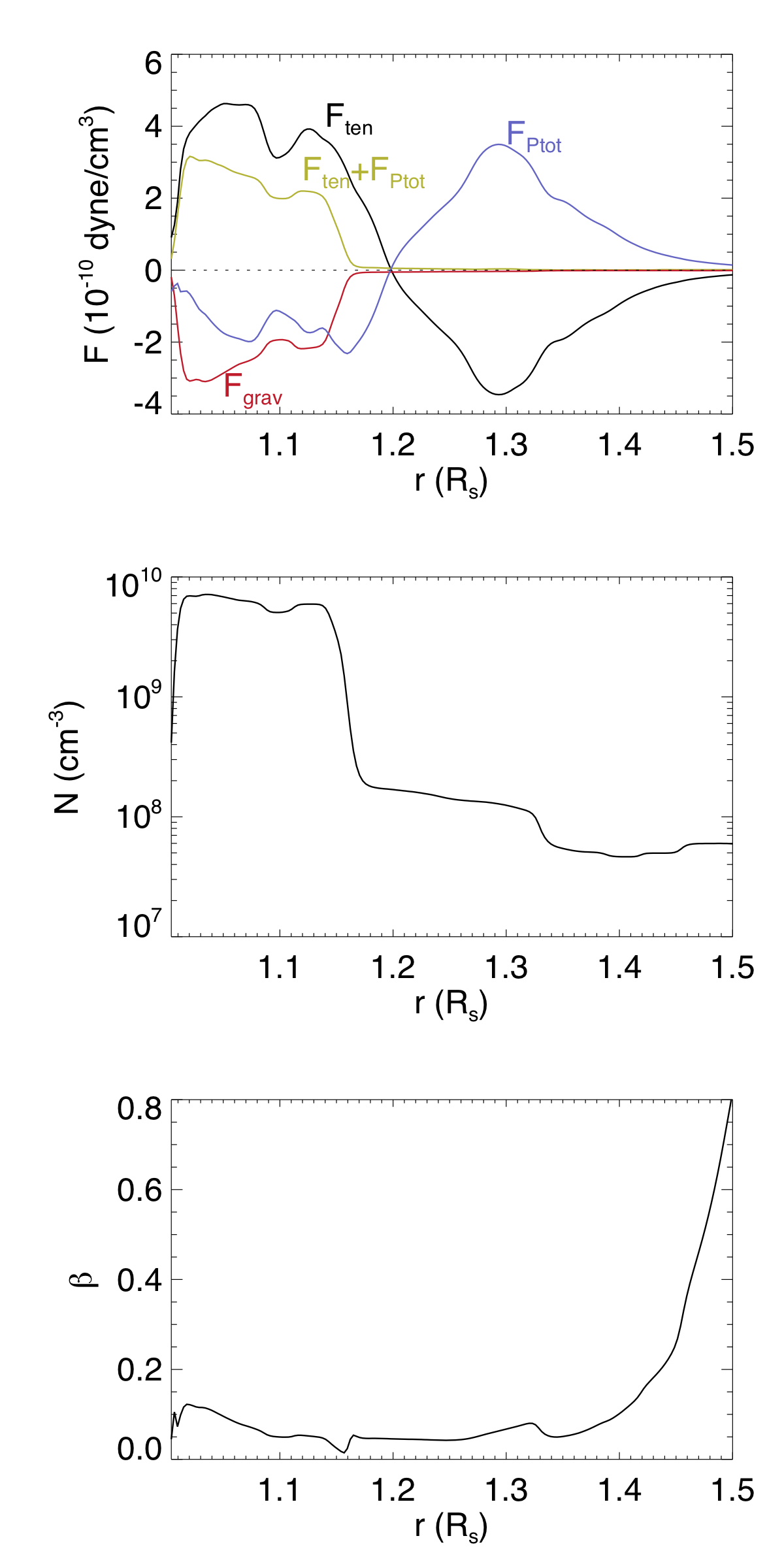}
\caption{Several radial forces (top), density (middle), and plasma-$\beta$
(the ratio of gas pressure over the magnetic pressure) (bottom)
along the central vertical line through the middle of the flux rope
shown in Figure \ref{fig:prom_evol}(b).
The radial forces shown in the top panel are the magnetic tension force
$F_{\rm ten}$ (black curve), the total pressure gradient force
$F_{\rm Ptot}$ (blue curve), for which the total pressure is
predominantly composed of the magnetic pressure (because of the low
plasma-$\beta$ as shown in the bottom panel),
the sum $F_{\rm ten}+F_{\rm Ptot}$ (green curve),
which is approximately the net Lorentz force,
and the gravity force of the plasma
$F_{\rm grav}$ (red curve).}
\label{fig:forces_prom}
\end{figure}

It has been shown for the WS-L case in F17 that the prominence carrying
magnetic field is significantly non-force-free.  This is further shown here
for the prominence carrying flux rope in the PROM simulation.
The top panel of Figure \ref{fig:forces_prom} shows several radial forces
along the central
vertical line through the middle of the flux rope shown in
Figure \ref{fig:prom_evol}(b) at time $t=17.25$ hours, during the
quasi-equilibrium phase.
The forces shown are the magnetic tension force
$F_{\rm ten}$ (black curve), the total pressure gradient force
$F_{\rm Ptot}$ (blue curve), for which the total pressure is
predominantly composed of the magnetic pressure 
because of the low plasma-$\beta$ as shown in the bottom panel,
the sum $F_{\rm ten}+F_{\rm Ptot}$ (green curve),
which is approximately the net Lorentz force,
and the gravity force of the plasma $F_{\rm grav}$ (red curve).
From the density profile in the middle panel, we can see that the height
range of the prominence (with density exceeding about
$5 \times 10^9 {\rm cm}^{-3}$)
extends from the base to
about $r=1.15 R_s$, which covers a broader range of heights when
compared to the WS-L case shown in Figure 19 of F17.
The top panel of Figure \ref{fig:forces_prom} shows that for the
height range containing the prominence,
there is a significant net upward Lorentz force (green curve)
that balances the downward gravity force
$F_{\rm grav}$ (red curve) of the prominence.
Thus the prominence carrying magnetic field is
significantly non-force-free (despite being low plasma-$\beta$),
with the downward prominence gravity and the downward magnetic
pressure gradient (blue curve) balancing the upward
magnetic tension. The prominence gravity accounts for a 
major portion of the magnetic tension force. In the rest of the
flux rope height range (from about $r=1.17 R_s$ to about $r=1.45 R_s$),
the net Lorentz force (green curve) is nearly zero, i.e. nearly
force-free, with the magnetic tension (black curve)
nearly balancing the magnetic pressure gradient (blue curve).
The significant gravity of the cool prominence mass on the 
magnetic field can suppress the kink instability of the
otherwise force-free flux rope.

\begin{figure}[hbt!]
\centering
\includegraphics[width=0.5\textwidth]{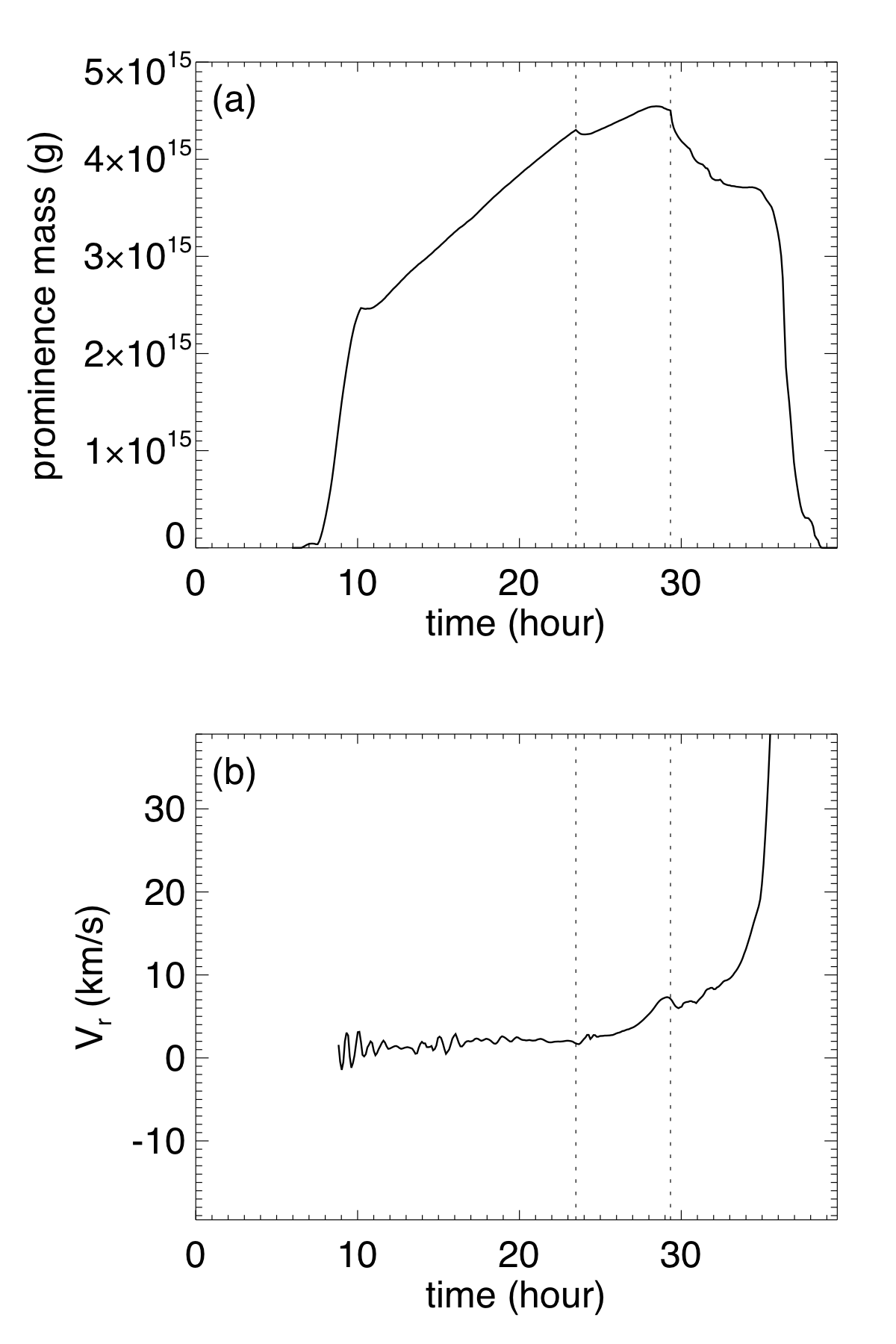}
\caption{(a) Temporal evolution of the cool prominence mass in the corona
evaluated as the total mass with temperature below $10^5$ K, and (b)
the same rise velocity of the flux rope as that shown
in Figure \ref{fig:emekvr}(b) but with a more zoomed-in view of the
evolution during the quasi-static phase for the PROM case.}
\label{fig:prommass_vrtracking}
\end{figure}

We find that the later onset of the kink motion of the flux
rope takes place when the prominence begins to show
draining and reduction of the prominence mass.
Figure \ref{fig:prommass_vrtracking}(a) shows the time variation of the total
cool prominence mass in the corona evaluated as the total mass
with temperature below $10^5$ K.  Figure \ref{fig:prommass_vrtracking}(b)
is the same rise velocity of the flux rope as that shown in
Figure \ref{fig:emekvr}(b) but shows a more zoomed-in view
of the evolution during the quasi-static phase.
The cool prominence mass shows mostly continuous increase in the earlier part
of the quasi-static phase. Later in the quasi-static phase we see episodes of
prominence mass draining, marked
in time by the two vertical dotted lines, corresponding to the time
instances shown in Figures \ref{fig:prom_evol}(i) and (j), where we see
branches of prominence draining developing (see also the online movie).
The latter monotonic decrease of the prominence
mass (following the time marked by the second dotted line)
corresponds with the final sustained acceleration and the dynamic eruption
of the flux rope.
Thus it appears that the initial growth of the prominence mass and its weight
is able to suppress the development of the kink instability over a long
period of time, from the time the emergence is stopped at $t=8.82$ hour
to at least about $t=23.5$ hour, for a period of about $110 \tau_A$,
during which the rise speed remains close to zero with no significant
acceleration (see Figure \ref{fig:prommass_vrtracking}(b)).
Some slow growth of the rise velocity and kink motion begin after
about $t=23.5$ hour (see Figure \ref{fig:prommass_vrtracking}(b) and
Figures \ref{fig:prom_evol}(c)(d)(e)),
when episodes of prominence draining develop that lighten the prominence
weight (Figures \ref{fig:prom_evol}(i)(j)(k) and
Figure \ref{fig:prommass_vrtracking}(a)).
The kink motion
in turn promotes the draining of the prominence mass because of upward
arching of the flux rope. Eventually the rise and
kink motion brings the central protrusion of the flux rope to a
height where it can no longer be confined and the flux rope undergoes an
ejective eruption as shown in Figure \ref{fig:prom_evol} and the on-line
movie.

\subsection{Formation of the prominence-cavity system}
\begin{figure}[hbt!]
\centering
\includegraphics[width=1.\textwidth]{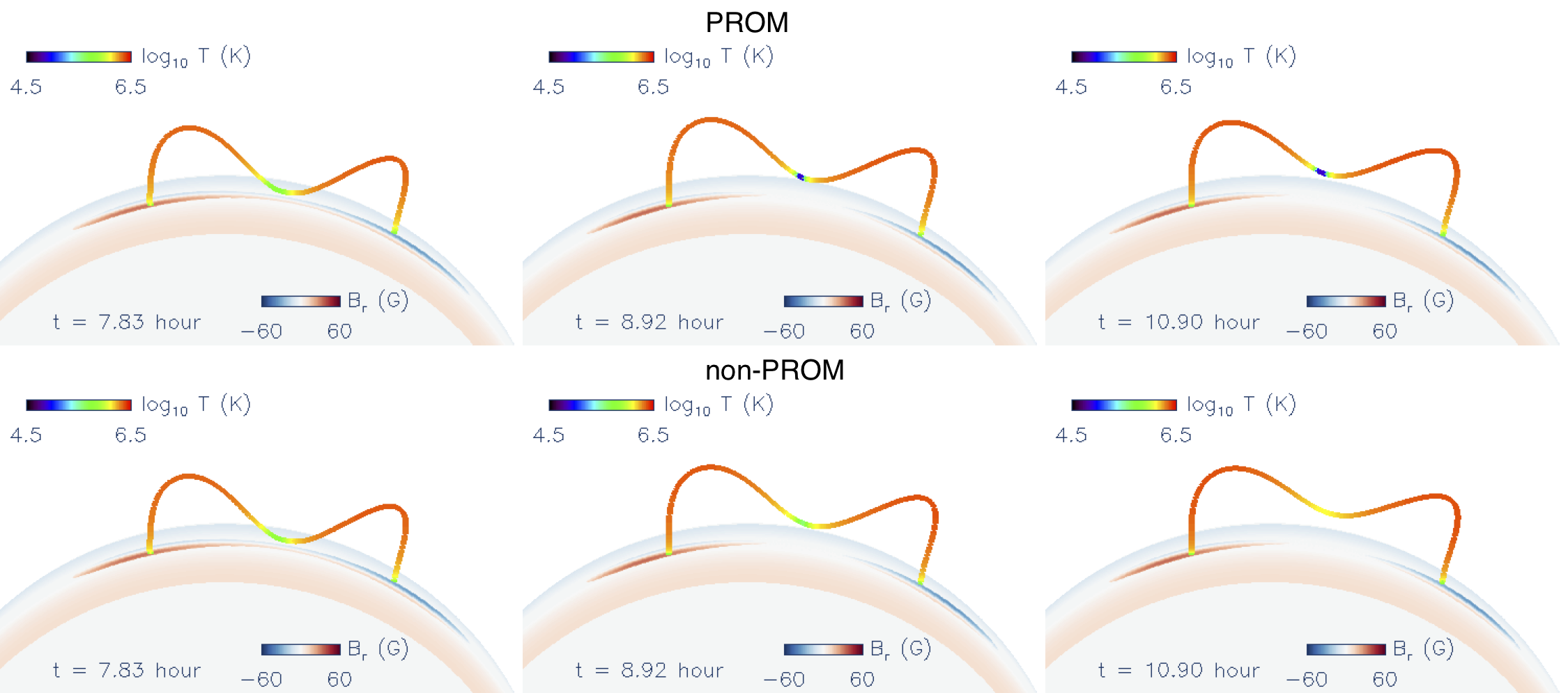}
\caption{Snapshots showing the evolution of a dipped field line colored with
temperature $T$. The upper row images show the PROM case and bottom row images
show the non-PROM case. The lower boundary sphere is colored with $B_r$.}
\label{fig:promfdl_temp}
\end{figure}

\begin{figure}[hbt!]
\centering
\includegraphics[width=1.\textwidth]{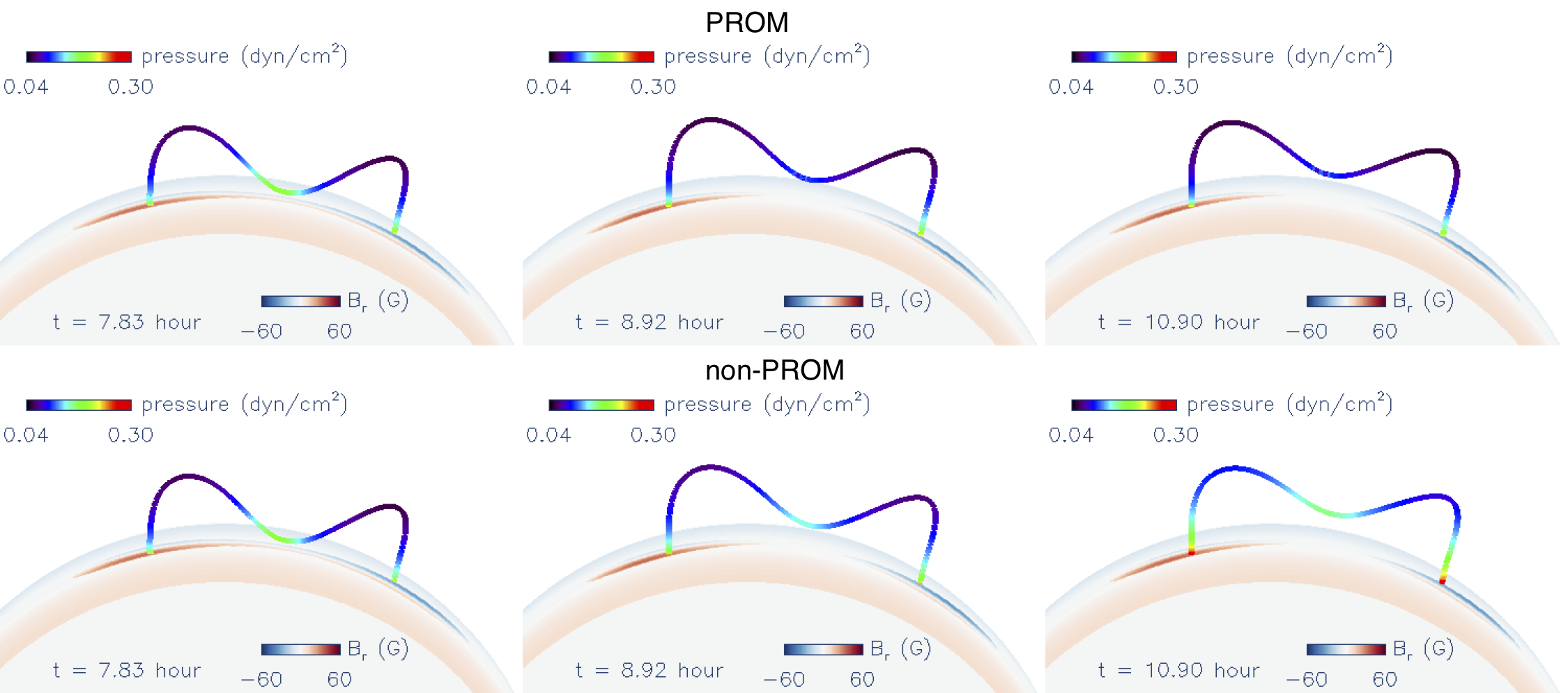}
\caption{Same as Figure \ref{fig:promfdl_temp} but with field line colored
with pressure.}
\label{fig:promfdl_pres}
\end{figure}

\begin{figure}[hbt!]
\centering
\includegraphics[width=1.\textwidth]{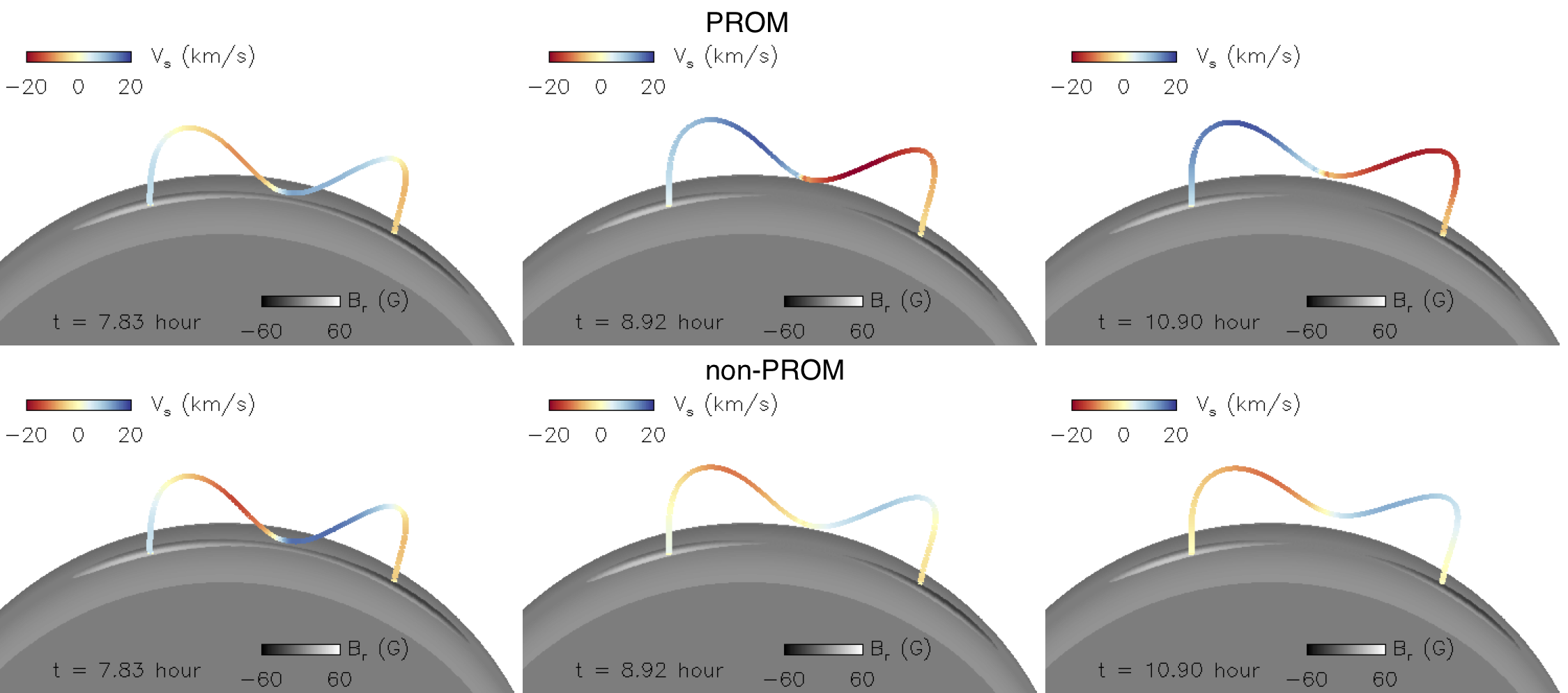}
\caption{Same as Figure \ref{fig:promfdl_temp} but with field line colored
with parallel velocity along the field line.}
\label{fig:promfdl_vs}
\end{figure}

\begin{figure}[hbt!]
\centering
\includegraphics[width=0.5\textwidth]{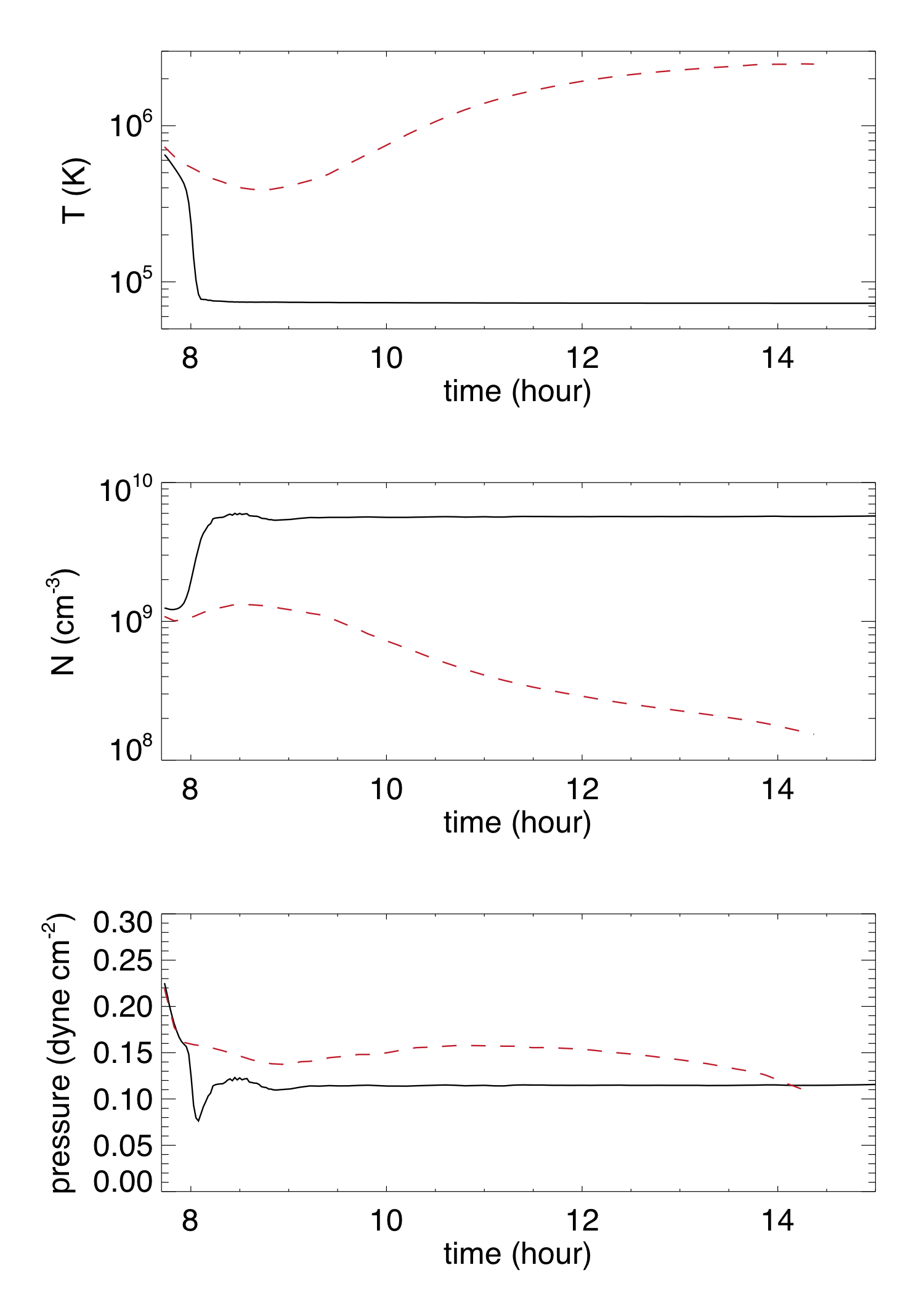}
\caption{Temporal evolution of the temperature (top), density (middle)
and pressure at the center of the dip for the PROM case (solid black curve)
and the non-PROM case (red dashed line).}
\label{fig:trackdip}
\end{figure}

Similar to the WS-L case shown in F17, we found in the PROM simulation that
prominence condensations form in the dips of the flux rope field lines 
due to the in-situ development of the radiative (thermal) instability
\citep[e.g.][]{Priest:2014,Xia:etal:2011} of the plasma in the dips
after their emergence into the corona.
Figures \ref{fig:promfdl_temp}, \ref{fig:promfdl_pres}, and \ref{fig:promfdl_vs}
show snapshots of a tracked field line that contains a dip, colored with 
temperature (Figure \ref{fig:promfdl_temp}),
pressure (Figure \ref{fig:promfdl_pres}),
and parallel velocity along the field line (Figure \ref{fig:promfdl_vs}),
for both the PROM case (upper row in the figures)
and the non-PROM case (bottom row in the figures).
Figure \ref{fig:trackdip} shows the temporal evolution of the
temperature (top panel),
density (middle panel), and pressure (bottom panel)
at the center of the dip for the PROM case
(black solid curve) and the non-PROM case (red dashed curve).
Just after
the emergence of the dip (see the left snapshots in Figures \ref{fig:promfdl_temp}
\ref{fig:promfdl_pres} and \ref{fig:promfdl_vs} at $t=7.83$ hours, and also
Figure \ref{fig:trackdip}), the emerged dip in both the PROM and
non-PROM case has a similar initial (coronal) temperature (about $7 \times 10^5$ K)
and density (about $10^9 \; {\rm cm}^{-3}$), with the dip in the PROM case being
slightly denser and cooler (Figure \ref{fig:trackdip}). Also initially
the parallel velocity shows an outflow from the dip in both the PROM and non-PROM
cases (left panels in Figure \ref{fig:promfdl_vs}).
Both the PROM dip and the non-PROM dip undergo an initial cooling and decrease in
pressure (see Figure \ref{fig:trackdip}), because they are not initially in
thermal equilibrium, until about $t=8$ hours, when the non-PROM dip
stabilizes to a relatively steady temperature (still coronal), density,
and pressure, but in contrast the
PROM dip undergoes a further, more rapid cooling, increase in density, and drop
in pressure. This divergence in behavior is mainly due to the different 
radiative cooling functions used for the two cases, where the PROM case undergoes
a run-away cooling because of the development of the radiative instability at the
dip, cooling all the way to the low temperature end (at about $7 \times 10^4$ K) of
the cooling function, forming a prominence condensation (see
the top row of Figure \ref{fig:promfdl_temp}, and the top panel in
Figure \ref{fig:trackdip}).
With the significantly larger decrease of pressure in the dip of the PROM case due
to prominence formation (see top panels of Figure
\ref{fig:promfdl_pres} and the solid curve in bottom panel of
Figure \ref{fig:trackdip}) compared to the non-PROM case (see the lower
panels of Figure \ref{fig:promfdl_pres} and the red dashed curve in bottom
panel of Figure \ref{fig:trackdip}), we find that 
the parallel velocity along the field line reverses direction to a converging
flow towards the dip in the PROM case
(see top panels of Figure \ref{fig:promfdl_vs}) while in
contrast the parallel velocity remains diverging away from the dip in the
non-PROM case (see bottom panels of Figure \ref{fig:promfdl_vs}).
Thus the prominence formation is driven at the dip due to in-situ development
of the radiative instability, similar to the case in F17, and the resulting
lowered pressure at the dip continues to draw inflow of mass towards the
dip from the footpoints (see the $t=8.92$ hours and $t=10.90$ hours panels 
in the top row of Figure \ref{fig:promfdl_vs}).
We find a continued inflow of mass towards the dip until about the time
the dip disappears.
With the use of a higher value of $C$ in equation (18) of F17 in the
current PROM simulation, which increases the coronal base pressure given the
same downward heat conduction flux, we find the formation of a more
massive and extended prominence with more dips
forming prominence condensations during the quasi-static phase compared to the WS-L case in F17.
This is because with a higher base pressure (and density), the density in the
emerged dips is higher, and thus more of the dips undergo the radiative
instability to form prominence condensations.
In Figure \ref{fig:prommass_vrtracking}(a) we see two distinct stages of
the prominence mass growth with different rates. One stage is from about
$t=8$ hours to about $t=10$ hours with a sharper growth, and switches
to a second stage of less steep growth from about $t=10$ hours to
$23.5$ hours. The initial faster growth stage is when the
newly emerged dips are undergoing radiative instability and forming
new prominence condensations. The second stage is after the emergence
has stopped, and no new prominence dips are forming, but the prominence
mass is still growing due to the continuing converging flows
into the dips that draw mass from the footpoints, as illustrated in the
example shown in the top right panel of Figure \ref{fig:promfdl_vs}.

\begin{figure}[hbt!]
\centering
\includegraphics[width=0.81\textwidth]{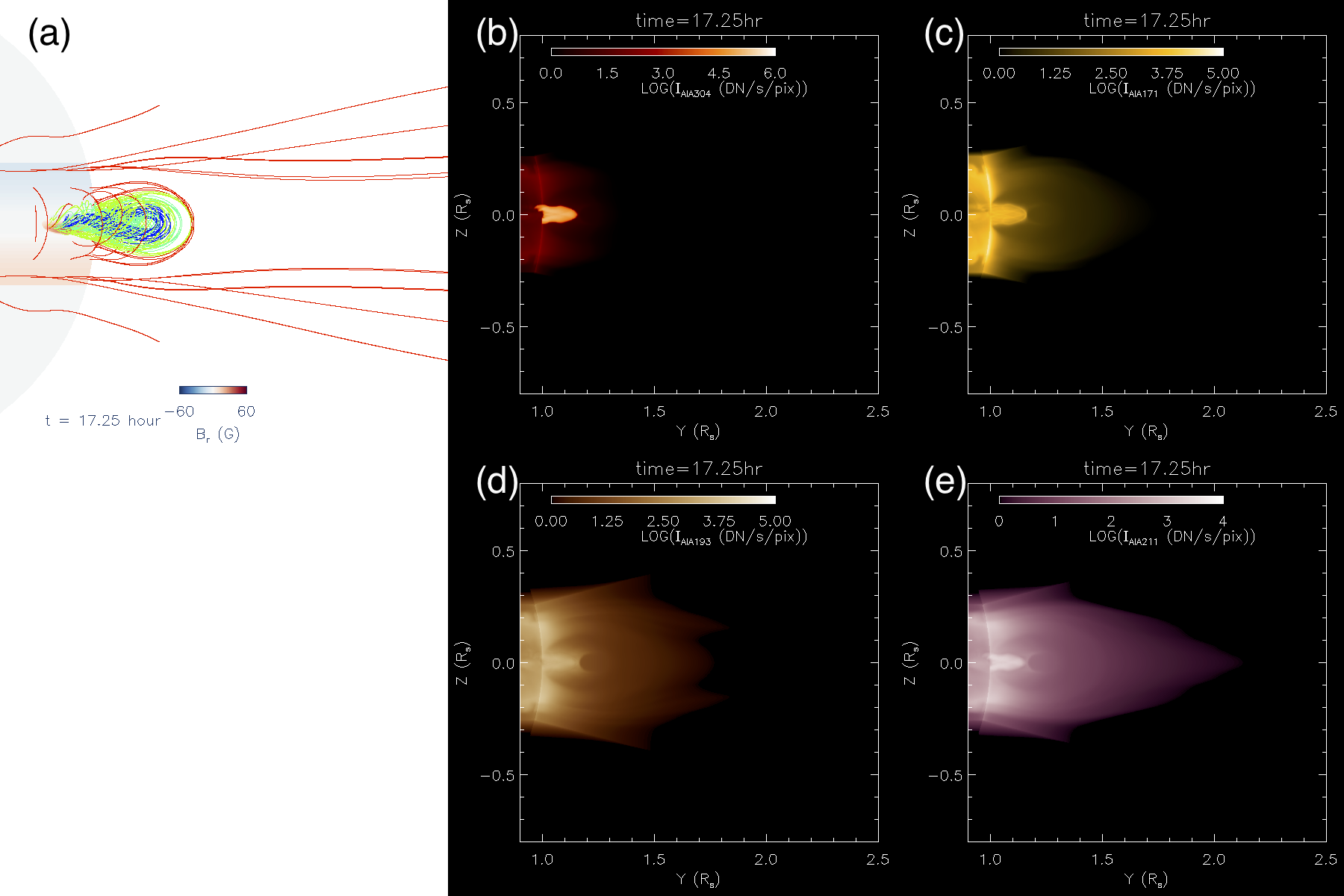}
\caption{(a) 3D view of the flux rope above
the limb viewed along its length, at the same time as that shown in
Figures \ref{fig:prom_evol}(b)(h),
and (b)-(e) the synthetic SDO/AIA images
in 304 {\AA}, 171{\AA}, 193{\AA}, 211{\AA} channel emission respectively,
viewed from the same LOS as that for (a).}
\label{fig:aia_prom_cavity_rot0}
\end{figure}

\begin{figure}[hbt!]
\centering
\includegraphics[width=0.81\textwidth]{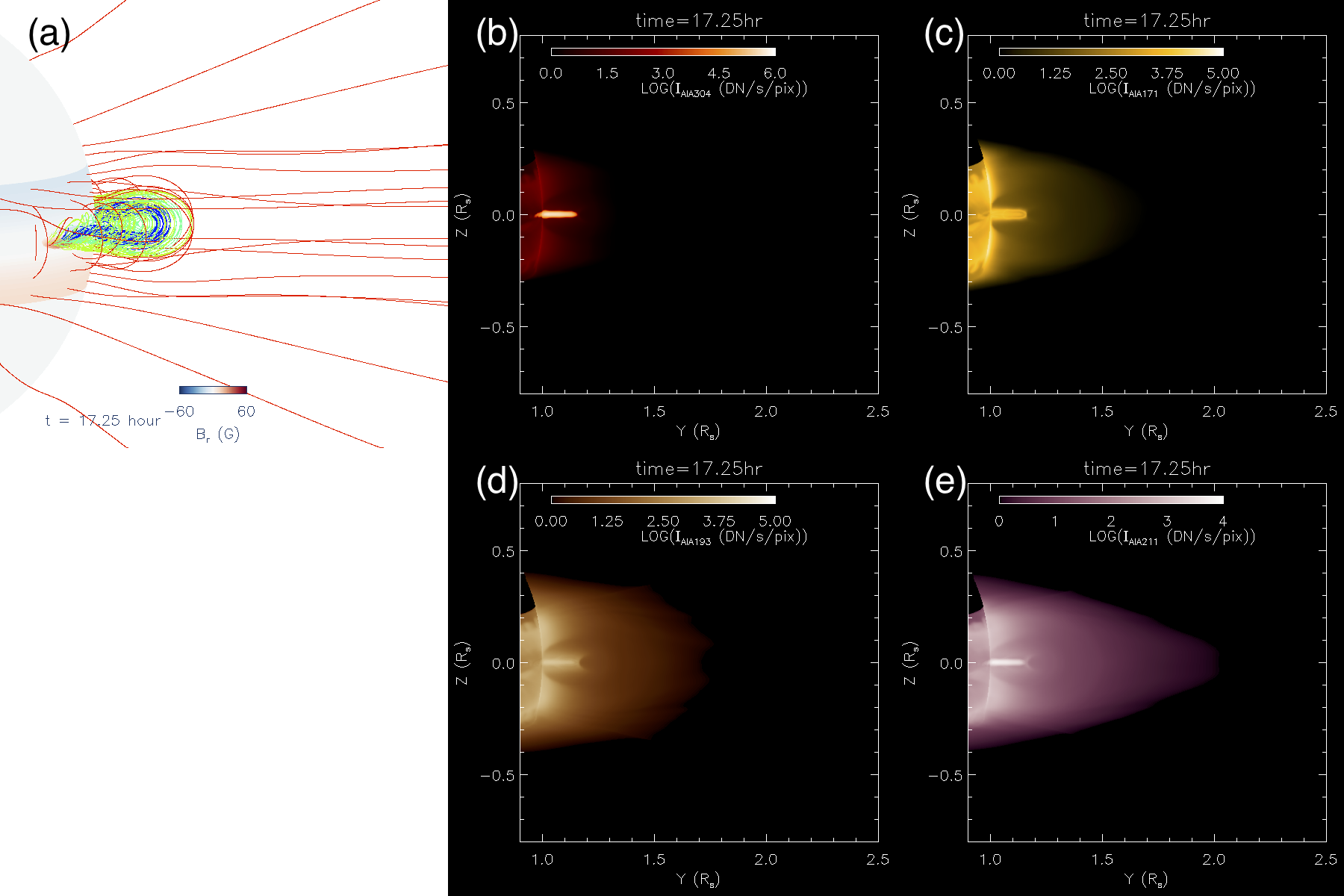}
\caption{Same as Figure \ref{fig:aia_prom_cavity_rot0} but viewed from a
slightly different LOS with the flux rope tilted anti-clockwise from the
longitudinal direction by $7^{\circ}$, such that the LOS is more aligned with
the prominence.}
\label{fig:aia_prom_cavity_rot7}
\end{figure}

\begin{figure}[htb!]
\centering
\includegraphics[width=0.37\textwidth]{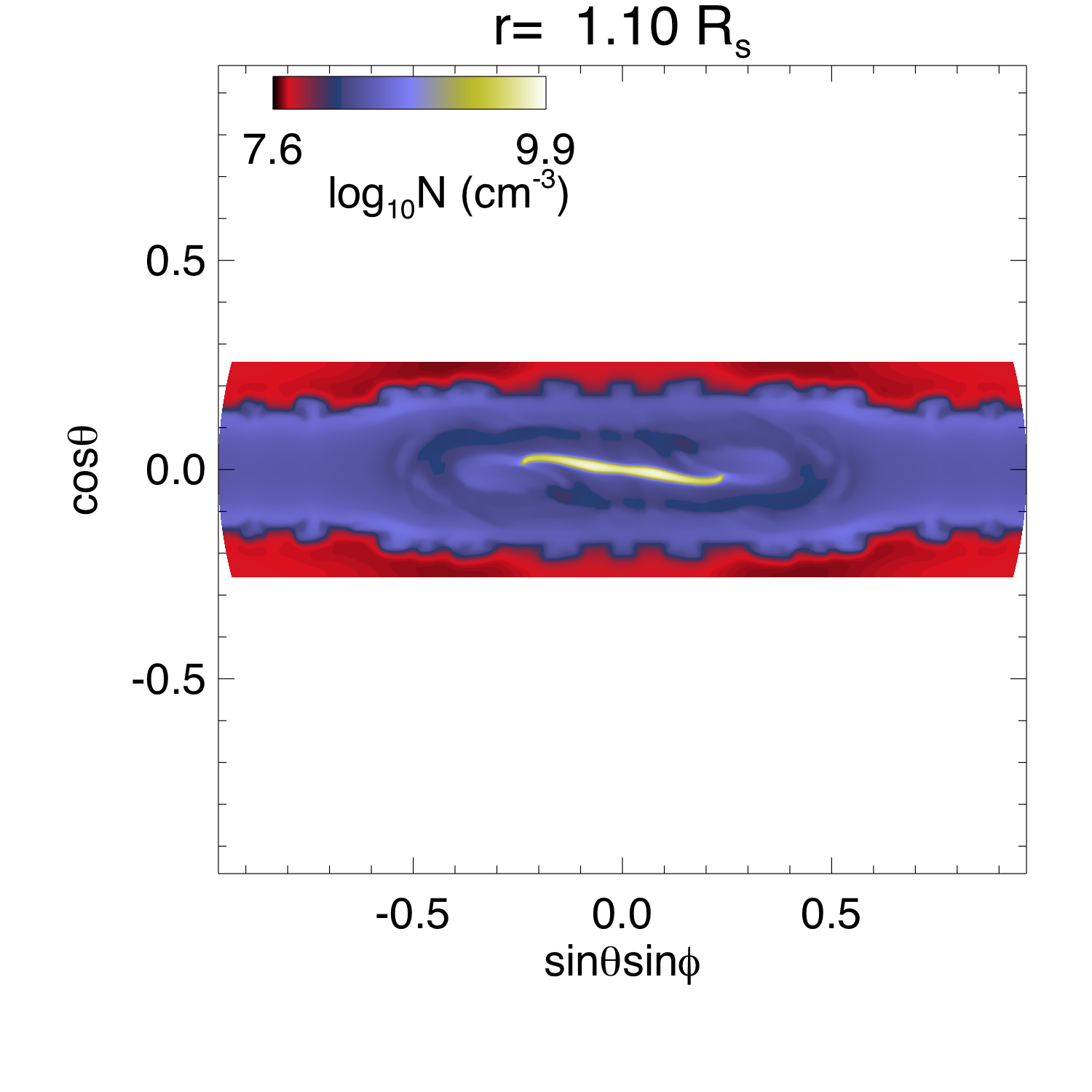}
\includegraphics[width=0.37\textwidth]{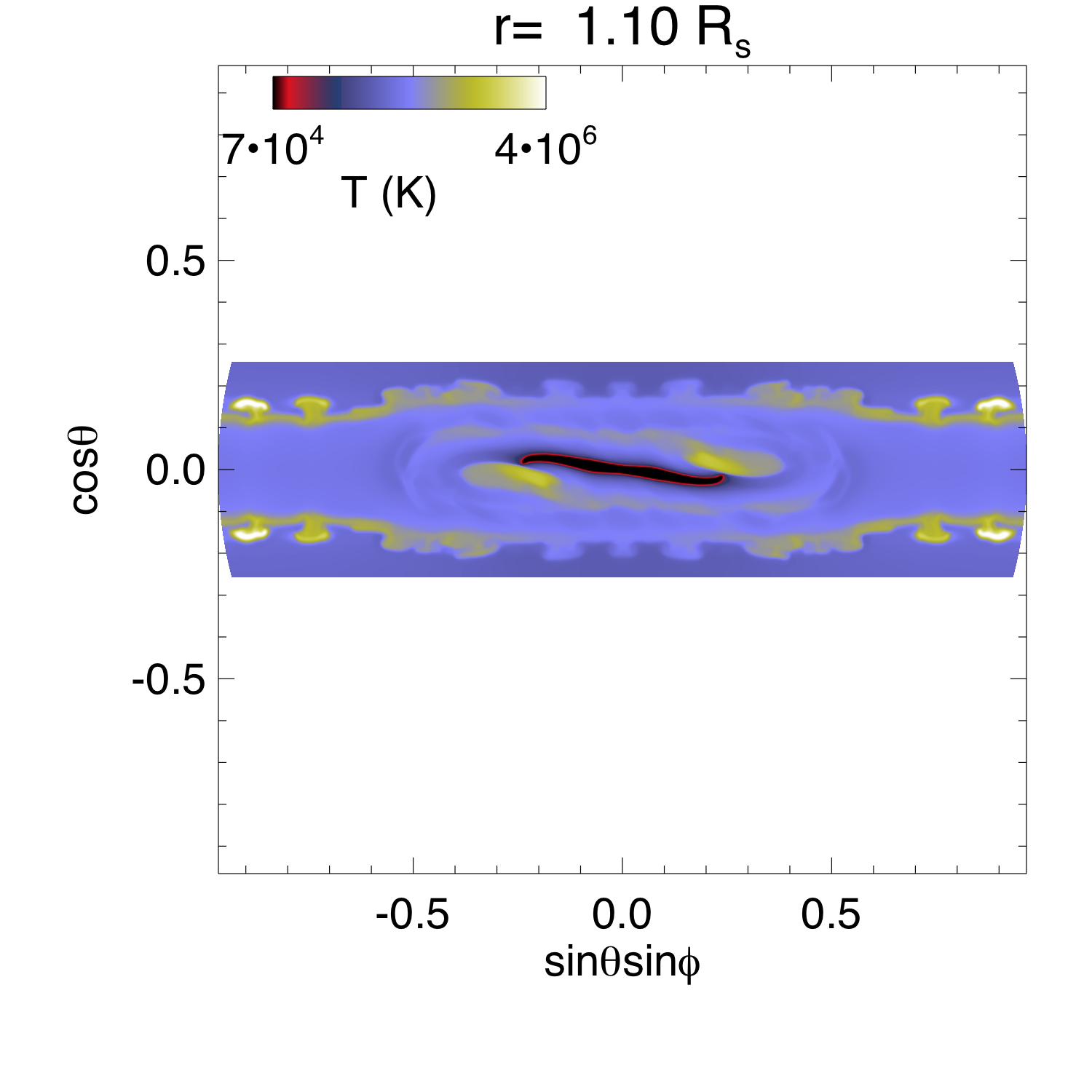}
\caption{Horizontal cross-section of density (left) and 
temperature (right) at the height of $r=1.1 R_s$, at time $t=17.25$ hour.}
\label{fig:rcut}
\end{figure}

\begin{figure}[hbt!]
\centering
\includegraphics[width=0.48\textwidth]{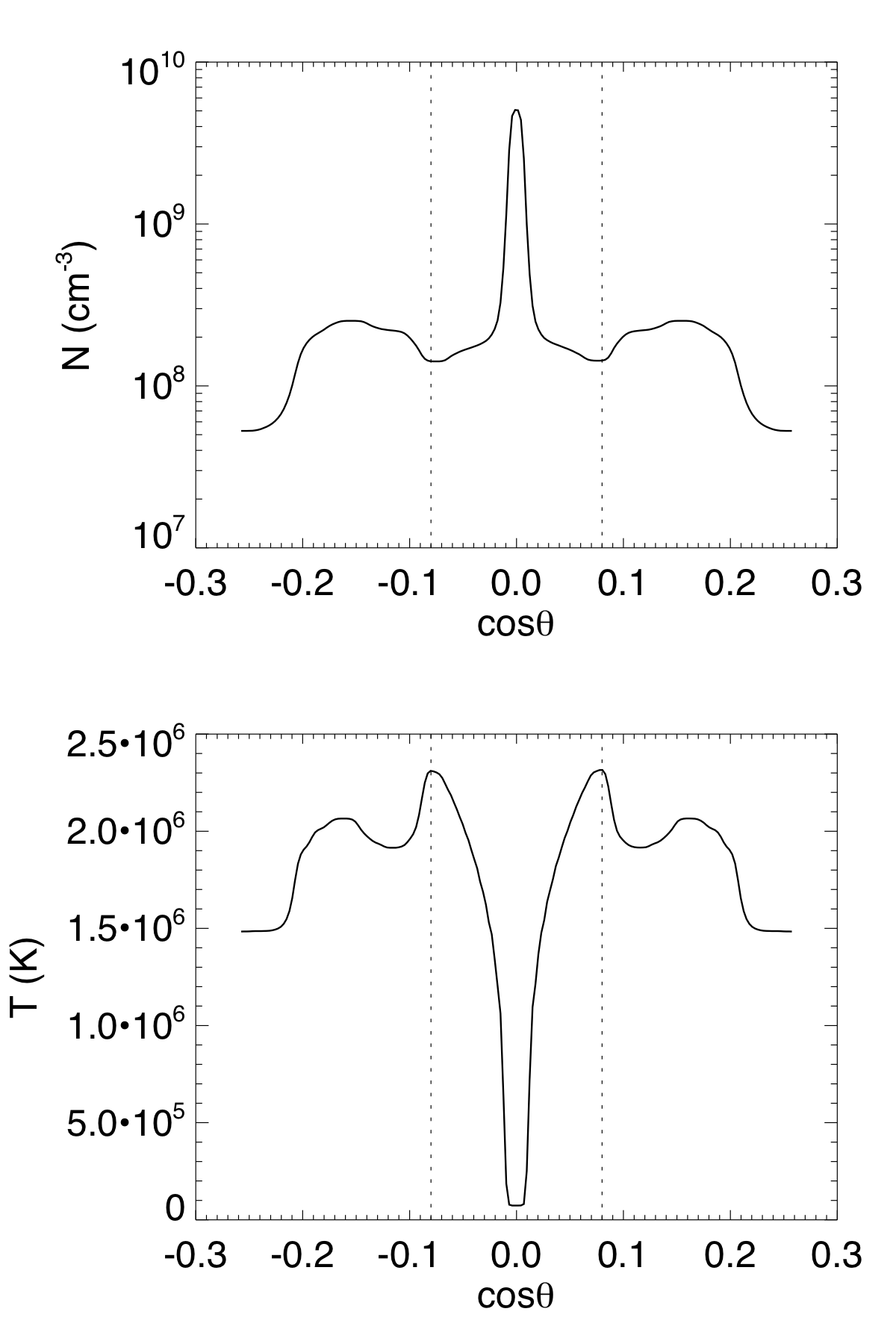}
\caption{The profile of density (top) and temperature (bottom) along the
central line at $\phi=0$ of the horizontal cross-section shown in
Figure \ref{fig:rcut}. The dotted lines mark the location of the cavity.}
\label{fig:prof_rcut}
\end{figure}

With the formation of a more extended prominence in the PROM simulation,
we also find the
formation of a cavity surrounding the prominence in the flux rope,
manifested in the synthetic AIA images shown in
Figures \ref{fig:aia_prom_cavity_rot0} and \ref{fig:aia_prom_cavity_rot7}
as viewed from two line-of-sights (LOS) that are nearly aligned with the
length of the flux rope, at time $t=17.25$ hours.
In the AIA 304 {\AA} images (Figures \ref{fig:aia_prom_cavity_rot0}(b)
and \ref{fig:aia_prom_cavity_rot7}(b)), we mainly see a narrow prominence
extending from the solar surface up to a height of about $0.15 R_s$ above
the limb. In the other hotter AIA channel images, we see that the bright narrow
prominence is surrounded by a relatively darker cavity region enclosed in the
brighter helmet dome, a configuration often seen in the observed coronal
prominence-cavity system \citep[e.g.][]{Regnier:etal:2011,Gibson:2015}.
The relatively darker cavity surrounding the prominence in the hotter
AIA channel images
(excluding the inner dark core seen on top of the prominence outlined
by bright ``horn"-like structures, to be discuss later)
is produce by the LOS integration along a lower density and higher
temperature tunnel that forms around the prominence.
Figure \ref{fig:rcut} shows a horizontal cross-section of
density (left panel) and
temperature (right panel) through the flux 
rope at the constant height of $r=1.1 R_s$, at the same time
as that shown in Figures \ref{fig:aia_prom_cavity_rot0}
and \ref{fig:aia_prom_cavity_rot7} at $t=17.25$ hours.
We see a central elongated filament condensation (of high density and
low temperature) with an inverse-S shape
surrounded by regions of relative low density and high temperature on both
sides of the filament.
Note that the rippling seen along the boundaries of the helmet fields
is due to the instability and reconnection intermittently taking
place higher up in different parts of the azimuthally extended current
sheet above the helmet streamer.
Figure \ref{fig:prof_rcut} shows the profile of density (top panel) and
temperature (bottom panel) along the slice through the center at $\phi=0$
of the horizontal cross-section shown in Figure \ref{fig:rcut}. 
It can be seen that the cavity regions (at the locations marked by the
dotted lines)
on two sides of the central filament condensation have a lower
density compared to the outer helmet rims and a relatively
hot temperature of about 2.3 MK.

\begin{figure}[htb!]
\centering
\includegraphics[width=0.8\textwidth]{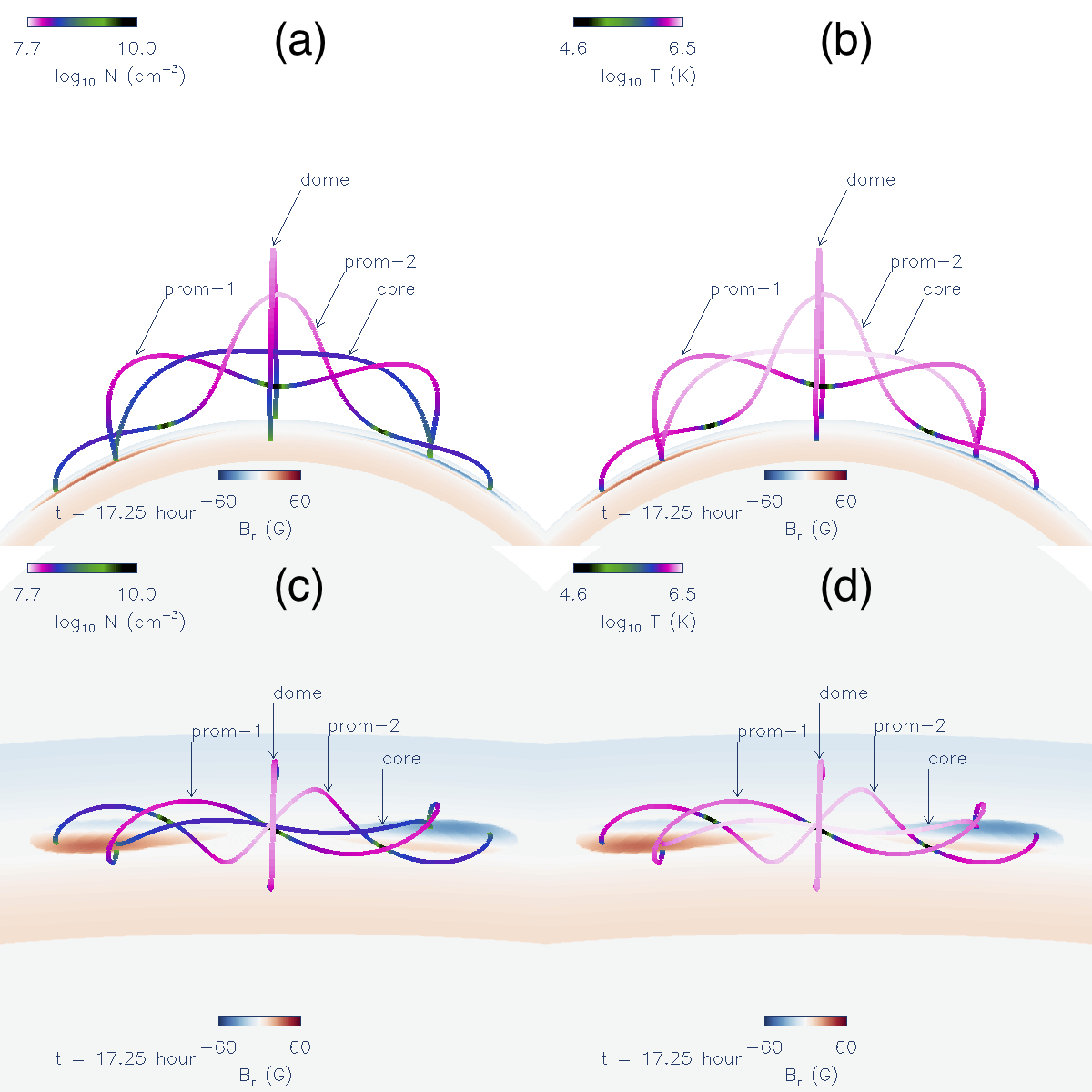}
\caption{A set of selected field lines colored according to density (left panels) and
temperature (right panels), as viewed from the side (upper rows) and
from above (bottom rows), at the same
time instance as that shown in Figure \ref{fig:prom_evol}(b)(h), during the
quasi-static phase.}
\label{fig:selected_fdls}
\end{figure}

\begin{figure}[htb!]
\centering
\includegraphics[width=0.5\textwidth]{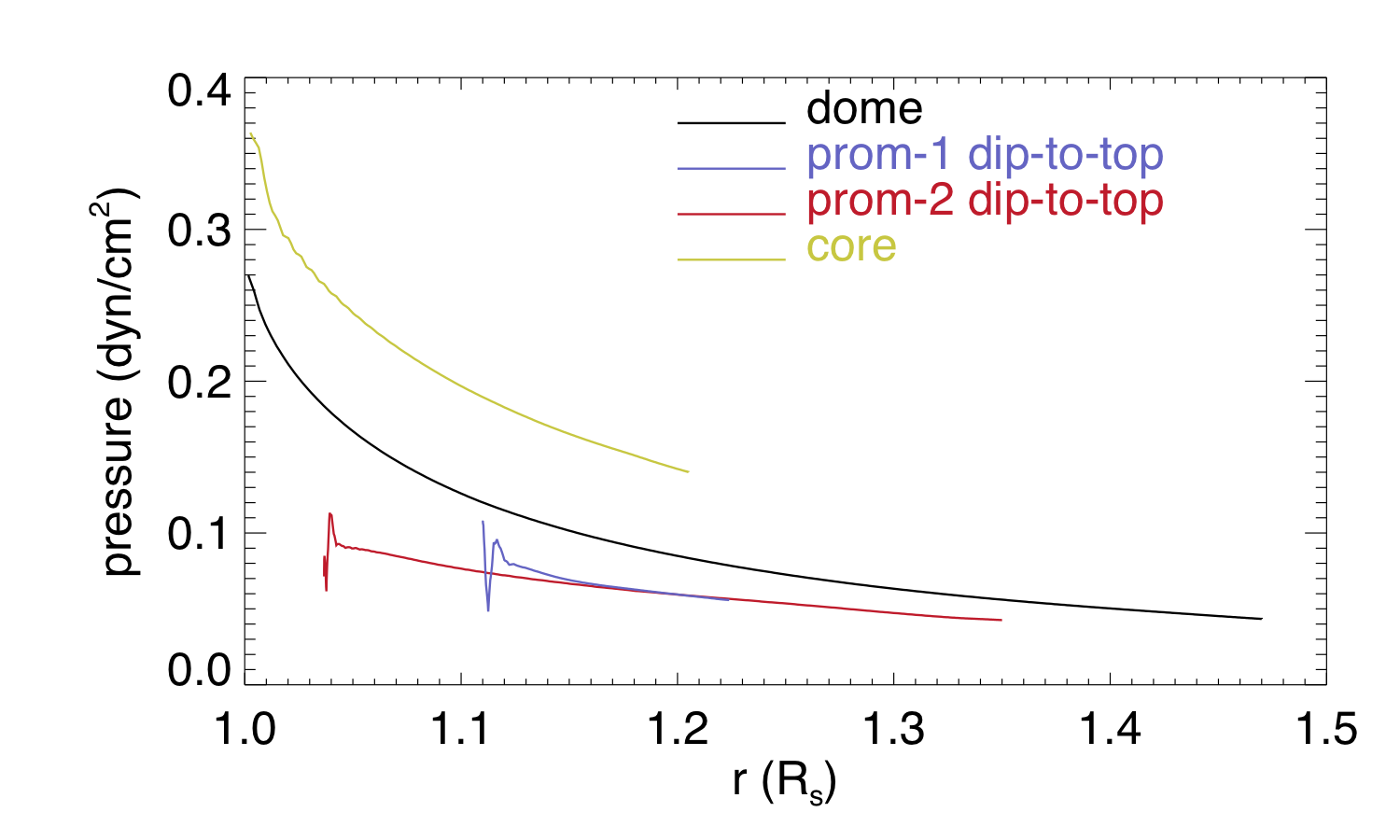} \\
\includegraphics[width=0.5\textwidth]{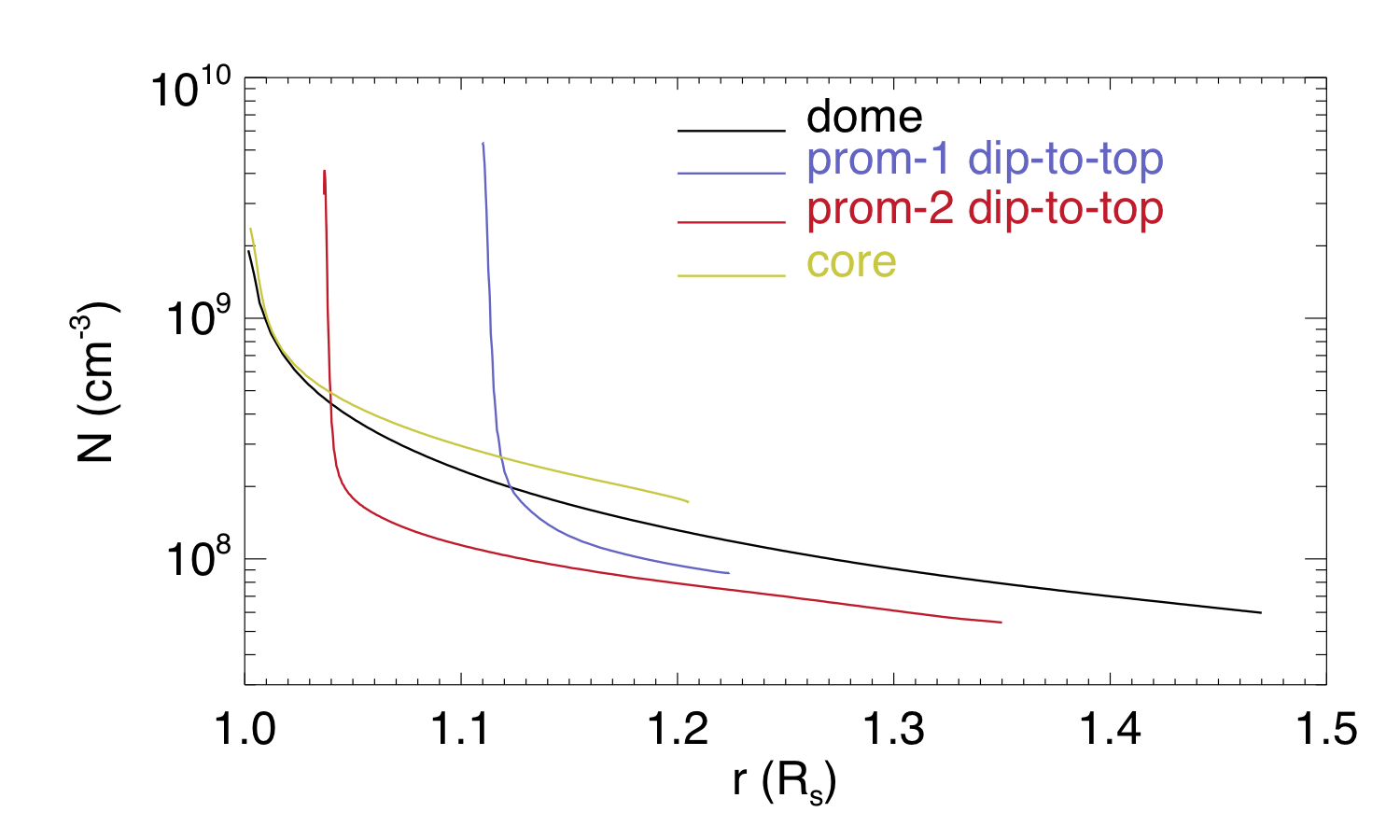} \\
\includegraphics[width=0.5\textwidth]{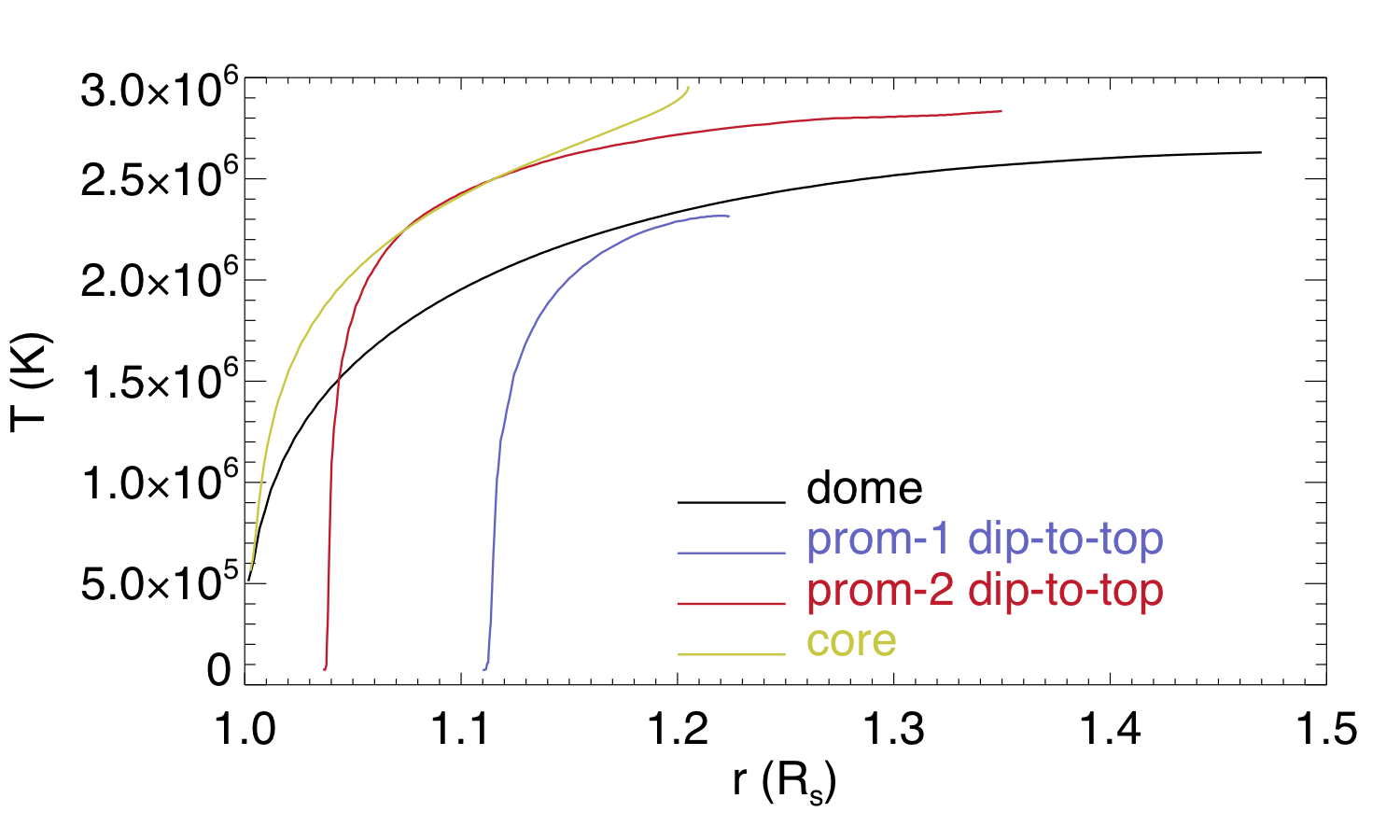}
\caption{The profiles of pressure (top), density (middle),
and temperature (bottom) as a function
of height along the ``dome'' field line (black curves) in 
Figure \ref{fig:selected_fdls}, and along the portion from the
bottom of the central dip to the right apex of the ``prom-1''
field line (blue curves) and the portion from the bottom of
the left dip to the central apex of the ``prom-2'' field line (red curves) in
Figure \ref{fig:selected_fdls}, and also along the ``core'' field line (gree curves)
in figure Figure \ref{fig:selected_fdls}.}
\label{fig:profs_fdls}
\end{figure}

To understand what types of the magnetic field make up the prominence and
the cavity, we show in Figure \ref{fig:selected_fdls}
a set of
representative field lines colored according to density (left panels) and
temperature (right panels),
viewed from two different perspectives (upper and lower rows).
Two of the field lines (labeled ``prom-1'' and ``prom-2'') are prominence
carrying field lines with
dips that contain prominence condensations (see the high density and
low temperature dips in Figure \ref{fig:selected_fdls}).
One of them (``prom-2'') has two prominence dips and
the other (``prom-1'') contains one prominence dip.
The field line labeled ``dome'' is a simple arcade field
line in the helmet dome.
The field line labeled ``core'' will be
discussed later.
It can be seen that for the prominence carrying field lines (``prom-1'' and ``prom-2''),
the portions extending up from the prominence dips have lower density than
the dome field line when compared at the same height (see Figure
\ref{fig:selected_fdls}(a)). This is most clearly seen by comparing the color
of the upper middle portion of the prom-2 field line with that of the
dome field line at the same height.
It can also be seen that the upper middle portion of
the prom-2 field line extending up from the dips shows higher temperature
than the dome field line when compared at the same height (see the field line
color in Figure \ref{fig:selected_fdls}(b)).
In Figure \ref{fig:profs_fdls} we plot the profiles of
pressure (top), density (middle), and temperature (bottom) as a function
of height along the dome field line (black curves),
the dip-to-top portions of the prom-1
(blue curves) and prom-2 (red curves) field lines, and the
core field line (green curves) shown in Figure
\ref{fig:selected_fdls}.
The blue curves in Figure \ref{fig:profs_fdls}
show the profiles along the portion of the prom-1 field line in
Figure \ref{fig:selected_fdls} from the bottom of the central dip to the right
apex, and the red curves in Figure \ref{fig:profs_fdls} show the profiles
along the portion of the prom-2 field line in Figure \ref{fig:selected_fdls}
from the bottom of the left dip to the central apex.
We find that the prominence carrying field lines develop a
lower pressure compared to the arcade dome field line at all heights (see
Figure \ref{fig:profs_fdls} top panel), as a result of the formation
of prominence condensations at the dips due to the run-away radiative cooling.
Note that although the mean pressure is lower at the dips, large
pressure fluctuations are seen at the bottom dips of the
prom-1 and prom-2 field lines along the nearly horizontal dipped
segments.
The cooling time scale near the dip can be as short as the
sound crossing time across the dip (F17) such that the hydrostatic balance
is not well maintained and there are large pressure fluctuations along
the dipped portion.
The reduced mean pressure at the dips due to prominence formation
drains plasma to the dips and establishes a more rarefied atmosphere
along the portions of the field lines extending above the dips,
resulting in lower density
in those portions of the prominence carrying field lines compared to
the dome field line at the same height (see Figure \ref{fig:profs_fdls} middle
panel). In some cases (such as for the prom-2 field line) it also leads
to higher temperature in the field line portions extending above the dips
compared to the dome field line (see the bottom panel of
Figure \ref{fig:profs_fdls}) because of the reduction of cooling by
the reduction of density.
The cavity is thus formed by the rarefied portions
of the prominence carrying field lines extending above the prominence dips.
We find in the PROM simulation that the presence of the type of field
lines containing two dips with prominence condensations
(e.g. the prom-2 field line)
in addition to the type (e.g. the prom-1 field line) that has only
one prominence dip, produces
a more continuous and extended low density cavity region along
the length of the flux rope compared to the WS-L case in F17, in which
only the latter type of prominence carrying field lines are present.
As a result in the PROM simulation we obtain a distinct cavity surrounding
the prominence in the AIA synthetic images viewed nearly along the length
of the flux rope (Figures
\ref{fig:aia_prom_cavity_rot0} and \ref{fig:aia_prom_cavity_rot7}).

However we note that within the cavity seen in the synthetic
AIA 193 {\AA} and 211 {\AA} channel images, there is a smaller
central dark core region on top of the prominence, outlined by
bright ``horn''-like structures. We found that this central dark core
is of a different character compared to the rest of the surrounding
dark cavity region that is formed by the LOS contribution from the
density depleted portions of the prominence carrying field lines extending
above the prominence dips as discussed above. We found that the
central dark core is threaded through by twisted field lines that
contain no dips, as represented
by the labeled ``core'' field line in Figure \ref{fig:selected_fdls},
which is hotter, has similar or higher density, and has higher
pressure compared to the dome field line at the same height
(see also the green curves showing the hight profiles of the
core field line in Figure \ref{fig:profs_fdls}). 
Thus the central dark core inside the cavity in the
AIA 193 {\AA} and 211 {\AA} channel images
is due to LOS contribution from highly heated,
twisted field lines that contain no dips, and it appears
dark not because of low density, but because its high
temperature (exceeding 3 MK) has gone above the peak
sensitivity temperatures of the AIA 193 {\AA} and
AIA 211 {\AA} channels. This cause is quite different from
that for the surrounding dark cavity region. We found that
the bright ``horn''-like structures extending from the prominence
is due to LOS contribution from field lines (not shown here) that
contain shallow dips that are about to disappear, and contain denser
plasma at coronal temperatures of about 1 to 2 MK that
the AIA 193 {\AA} and AIA 211 {\AA} channels are sensitive to.
Internal features inside a prominence cavity with a central
dark region outlined by bright U-shaped prominence ``horns'' on top of
the prominence has been observed in AIA 171 {\AA}, 193 {\AA}, and
211 {\AA} channel images of prominence cavities
\citep[e.g.][]{Berger:2012,Su:etal:2015}.
The prominence-cavity system with the above internal
features has also been produced
by previous simulations of prominence formation by
\citet{Xia:etal:2014} and \citet{Xia:Keppens:2016a},
which include the chromosphere at
the lower boundary. Their interpretation of the structures obtained
from the simulation is different from the nature of the structures
found in our simulation.
They found that the central dark cavity
enclosed by the horns is threaded by both prominence-free dipped
field lines and arched twisted field lines with no dips, while the
outer cavity is formed by arched twisted field lines with no dips
\citep{Xia:etal:2014}. They found that during prominence-cavity
formation, density depletion occurs not only on prominence-loaded field
lines threading cavity and prominence where in situ condensation happens
(as is the case in our simulation),
but also on prominence-free field lines due to mass drainage into the
chromosphere. Their model of prominence-cavity formation is likely
more realistic because of the inclusion of the chromosphere, and it
allows the formation of filament channels or coronal cavities
in absence of filament or prominence condensations.

\section{Conclusions}
Using the same numerical model described in F17, we have carried out
simulations of the evolution of a coronal flux rope emerging into a
coronal streamer, studying the cases with and without the formation
of prominence condensations in the dips of the twisted flux rope field lines.
In the case with prominence formation (the PROM simulation), we
obtained the formation of a more massive and extended prominence in
the emerged flux rope compared to the similar simulation in F17 (the WS-L
simulation in that paper) by changing the specification of the lower
boundary coronal base pressure.
We found prominence condensations forming in
more dips of the twisted field lines by the development of
the radiative instability, and a more continuous and extended
low density cavity region surrounding the elongated prominence/filament 
in the flux rope.
The elongated filament is found to display a sigmoid shape.
Also similar to the observations of prominence-cavity systems,
a distinct dark cavity can be seen around
the central narrow prominence in the helmet dome in the
synthetic AIA 171 {\AA}, 193 {\AA}, and 211 {\AA} emission images
with the flux rope above the limb viewed along LOSs nearly
aligned with its length.
It is found that the cavity region is formed by the portions of the
prominence carrying field lines that extend upward from the prominence
dips, forming a more rarefied atmosphere along these portions of the field
lines (compared to the dome arcade field lines) due to
the lowered pressure at the dips by the formation of prominence
condensations.  The cavity region is also found to develop hotter
temperatures (2 - 3 MK ) compared to the rim of the helmet dome (see
e.g. red curve in bottom panel of Figure \ref{fig:profs_fdls}).
In addition, it is found that within the cavity seen in the
synthetic AIA 193 {\AA} and 211 {\AA} channel images, there is an apparent inner dark
``cavity'' on top of the prominence, outlined by bright ``horn''-like
structures. In our PROM simulation, this inner dark ``cavity'' however is
of a different nature compared to the rest of the surrounding dark cavity: it is
produced by LOS contribution from highly heated, twisted field lines 
that contain no dips, and appears dark not because of low density, but
because of the high temperature (exceeding 3 MK) that has gone
significantly above the peak sensitivity temperatures of the AIA 193 
{\AA} and AIA 211 {\AA} channels.

Similar to F17, we found that the magnetic field in the region of the
prominence condensations is significantly non-force-free, despite being
low plasma-$\beta$.  A significant non-zero net Lorentz force,
corresponding to a major fraction of the upward magnetic tension force
of the dipped field lines, balances the weight of the cool prominence
mass. Thus the prominence mass is playing a significant role in the
stability and confinement of the coronal flux rope.
By comparing the simulations with
and without the formation of prominence condensations, we found that
the weight of the prominence mass can suppress or delay the onset of
the kink instability of an otherwise nearly force-free flux rope. It
allows the flux rope in the PROM simulation to be confined
in equilibrium significantly longer compared to the non-PROM case.
The kink motion and the eruption eventually develop for the prominence
carrying flux rope with substantial draining of the prominence plasma,
which lightens the weight on the magnetic field.
Our simulations suggest that the prominence mass is dynamically important
for the equilibrium buildup and eruptive properties of CMEs.
During the onset of the eruption, the synthetic AIA 304 images show that
the prominence is lifted up into an erupting loop with helical features along
the loop and significant draining along the legs of the loop, as often
seen in observations.

\acknowledgments
I thank Joan Burkepile for reading and helpful comments on the paper.
I also thank the anonymous referee for helpful comments that improved the paper.
This work is supported in part by the Air Force Office of Scientific Research
grant FA9550-15-1-0030 to NCAR. NCAR is sponsored by the National Science
Foundation. The numerical simulations were carried out on the Cheyenne
supercomputer at NWSC under the NCAR Strategic capability project NHAO0011
and also on the DOD supercomputer Topaz at ERDC under the
project AFOSR4033B701.


\clearpage
\begin{thebibliography}{}
\expandafter\ifx\csname natexlab\endcsname\relax\def\natexlab#1{#1}\fi
\providecommand{\url}[1]{\href{#1}{#1}}

\bibitem[{{Amari} {et~al.}(2014){Amari}, {Canou}, \& {Aly}}]{Amari:etal:2014}
{Amari}, T., {Canou}, A., \& {Aly}, J.-J. 2014, \nat, 514, 465

\bibitem[{{Athay}(1986)}]{Athay:1986}
{Athay}, R.~G. 1986, \apj, 308, 975

\bibitem[{{Aulanier} {et~al.}(2010){Aulanier}, {T{\"o}r{\"o}k}, {D{\'e}moulin},
  \& {DeLuca}}]{Aulanier:etal:2010}
{Aulanier}, G., {T{\"o}r{\"o}k}, T., {D{\'e}moulin}, P., \& {DeLuca}, E.~E.
  2010, \apj, 708, 314

\bibitem[{{Berger}(2012)}]{Berger:2012}
{Berger}, T. 2012, in Astronomical Society of the Pacific Conference Series,
  Vol. 463, Second ATST-EAST Meeting: Magnetic Fields from the Photosphere to
  the Corona., ed. T.~R. {Rimmele}, A.~{Tritschler}, F.~{W{\"o}ger},
  M.~{Collados Vera}, H.~{Socas-Navarro}, R.~{Schlichenmaier}, M.~{Carlsson},
  T.~{Berger}, A.~{Cadavid}, P.~R. {Gilbert}, P.~R. {Goode}, \&
  M.~{Kn{\"o}lker}, 147

\bibitem[{{Chatterjee} \& {Fan}(2013)}]{Chatterjee:Fan:2013}
{Chatterjee}, P., \& {Fan}, Y. 2013, \apjl, 778, L8

\bibitem[{{Downs} {et~al.}(2012){Downs}, {Roussev}, {van der Holst}, {Lugaz},
  \& {Sokolov}}]{Downs:etal:2012}
{Downs}, C., {Roussev}, I.~I., {van der Holst}, B., {Lugaz}, N., \& {Sokolov},
  I.~V. 2012, \apj, 750, 134

\bibitem[{{Fan}(2010)}]{Fan:2010}
{Fan}, Y. 2010, \apj, 719, 728

\bibitem[{{Fan}(2012)}]{Fan:2012}
---. 2012, \apj, 758, 60

\bibitem[{{Fan}(2017)}]{Fan:2017}
---. 2017, \apj, 844, 26

\bibitem[{{Fan} \& {Gibson}(2007)}]{Fan:Gibson:2007}
{Fan}, Y., \& {Gibson}, S.~E. 2007, \apj, 668, 1232

\bibitem[{{Gibson}(2015)}]{Gibson:2015}
{Gibson}, S. 2015, in Astrophysics and Space Science Library, Vol. 415, Solar
  Prominences, ed. J.-C. {Vial} \& O.~{Engvold}, 323

\bibitem[{{Hood} \& {Priest}(1981)}]{Hood:Priest:1981}
{Hood}, A.~W., \& {Priest}, E.~R. 1981, Geophysical and Astrophysical Fluid
  Dynamics, 17, 297

\bibitem[{{Jenkins} {et~al.}(2018){Jenkins}, {Long}, {van Driel-Gesztelyi}, \&
  {Carlyle}}]{Jenkins:etal:2017}
{Jenkins}, J.~M., {Long}, D.~M., {van Driel-Gesztelyi}, L., \& {Carlyle}, J.
  2018, \solphys, 293, 7

\bibitem[{{Low}(1996)}]{Low:1996}
{Low}, B.~C. 1996, \solphys, 167, 217

\bibitem[{{Low}(2001)}]{Low:2001}
---. 2001, \jgr, 106, 25141

\bibitem[{{Munro} {et~al.}(1979){Munro}, {Gosling}, {Hildner}, {MacQueen},
  {Poland}, \& {Ross}}]{Munro:etal:1979}
{Munro}, R.~H., {Gosling}, J.~T., {Hildner}, E., {et~al.} 1979, Sol. Phys., 61,
  201

\bibitem[{{Priest}(2014)}]{Priest:2014}
{Priest}, E. 2014, {Magnetohydrodynamics of the Sun} (Cambridge, UK: Cambridge
  University Press)

\bibitem[{{R{\'e}gnier} {et~al.}(2011){R{\'e}gnier}, {Walsh}, \&
  {Alexander}}]{Regnier:etal:2011}
{R{\'e}gnier}, S., {Walsh}, R.~W., \& {Alexander}, C.~E. 2011, \aap, 533, L1

\bibitem[{{Su} {et~al.}(2015){Su}, {van Ballegooijen}, {McCauley}, {Ji},
  {Reeves}, \& {DeLuca}}]{Su:etal:2015}
{Su}, Y., {van Ballegooijen}, A., {McCauley}, P., {et~al.} 2015, \apj, 807, 144

\bibitem[{{T{\"o}r{\"o}k} \& {Kliem}(2005)}]{Toeroek:Kliem:2005}
{T{\"o}r{\"o}k}, T., \& {Kliem}, B. 2005, \apjl, 630, L97

\bibitem[{{T{\"o}r{\"o}k} \& {Kliem}(2007)}]{Toeroek:Kliem:2007}
---. 2007, Astronomische Nachrichten, 328, 743

\bibitem[{{T{\"o}r{\"o}k} {et~al.}(2011){T{\"o}r{\"o}k}, {Panasenco}, {Titov},
  {Miki{\'c}}, {Reeves}, {Velli}, {Linker}, \& {De Toma}}]{Toeroek:etal:2011}
{T{\"o}r{\"o}k}, T., {Panasenco}, O., {Titov}, V.~S., {et~al.} 2011, \apjl,
  739, L63

\bibitem[{{T{\"o}r{\"o}k} {et~al.}(2018){T{\"o}r{\"o}k}, {Downs}, {Linker},
  {Lionello}, {Titov}, {Miki{\'c}}, {Riley}, {Caplan}, \&
  {Wijaya}}]{Toeroek:etal:2018}
{T{\"o}r{\"o}k}, T., {Downs}, C., {Linker}, J.~A., {et~al.} 2018, ArXiv
  e-prints, arXiv:1801.05903

\bibitem[{{Webb} \& {Hundhausen}(1987)}]{Webb:Hundhausen:1987}
{Webb}, D.~F., \& {Hundhausen}, A.~J. 1987, \solphys, 108, 383

\bibitem[{{Xia} {et~al.}(2011){Xia}, {Chen}, {Keppens}, \& {van
  Marle}}]{Xia:etal:2011}
{Xia}, C., {Chen}, P.~F., {Keppens}, R., \& {van Marle}, A.~J. 2011, \apj, 737,
  27

\bibitem[{{Xia} \& {Keppens}(2016)}]{Xia:Keppens:2016a}
{Xia}, C., \& {Keppens}, R. 2016, \apj, 823, 22

\bibitem[{{Xia} {et~al.}(2014){Xia}, {Keppens}, {Antolin}, \&
  {Porth}}]{Xia:etal:2014}
{Xia}, C., {Keppens}, R., {Antolin}, P., \& {Porth}, O. 2014, \apjl, 792, L38

\end{thebibliography}
\end{document}